\documentclass[pdfoutput=1]{statsoc}

\usepackage[left=7mm,right=15mm]{geometry} 
\usepackage{vmargin} 
\usepackage{hyperref} 

\usepackage[english]{babel}
\usepackage{amsmath,amssymb,amsfonts,latexsym,natbib,url}
\usepackage[usenames]{color}
\usepackage[utf8]{inputenc}
\usepackage{graphicx,siunitx}
\usepackage{tikz}
\usetikzlibrary{arrows,chains,matrix,positioning,scopes,shapes,automata,mindmap,snakes,shadows}

\makeatletter
\tikzset{join/.code=\tikzset{after node path={%
\ifx\tikzchainprevious\pgfutil@empty\else(\tikzchainprevious)%
edge[every join]#1(\tikzchaincurrent)\fi}}}
\makeatother
\tikzset{>=stealth',every on chain/.append style={join},
         every join/.style={->}}
\tikzstyle{labeled}=[execute at begin node=$\scriptstyle,
   execute at end node=$]

\newtheorem{prop}{Proposition}

\newtheorem{hyp}{Assumption}

\newtheorem{remark}{Remark}
\newtheorem{example}{Example}

\newcommand{\R}{\mathbb{R}} 
\newcommand{\N}{\mathbb{N}} 

\newcommand{\pr}{\mathbb{P}_{\theta}}
 
\newcommand{\qr}{\mathbb{Q}} 
\newcommand{\esp}{\mathbb{E}}
\newcommand{\h}{\mathcal{H}} 
 
\newcommand{\argmax}{\textrm{Argmax}}

\newcommand{\Q}{\mathcal{Q}} 
\newcommand{\Y}{\mathbf{Y}} 
\newcommand{\Z}{\mathbf{Z}} 
 
\newcommand{\bpi}{\boldsymbol{\pi}}
\newcommand{\ba}{\boldsymbol{\alpha}}
\newcommand{\bb}{\boldsymbol{\beta}}
\newcommand{\bg}{\boldsymbol{\gamma}}
\newcommand{\taum}{\tau_{\text{marg}}}
\newcommand{\htaum}{\hat{\tau}_{\text{marg}}}

\title[Clustering dynamic random graphs via SBM]{Statistical clustering of temporal networks through a dynamic 
  stochastic block model}
\author[C. Matias and V. Miele]{Catherine Matias\footnote{Corresponding author.}}
\address{
Sorbonne Universités, UPMC Univ Paris 06, Univ Paris Diderot, Sorbonne Paris Cité, 
CNRS, Laboratoire de Probabilités et Modèles Aléatoires (LPMA), 75005 Paris, France.}
\email{catherine.matias@math.cnrs.fr}
\author[C. Matias and V. Miele]{Vincent Miele} 
\address{Université de Lyon, F-69000 Lyon; Université Lyon 1; CNRS, UMR5558, \\
Laboratoire de Biométrie et Biologie \'Evolutive,
F-69622 Villeurbanne, France. }
\email{vincent.miele@univ-lyon1.fr}

\begin{document}

 \begin{abstract}
Statistical node clustering in discrete time dynamic networks is an emerging field that raises many challenges. Here, we explore statistical
properties and frequentist inference in a model that combines a stochastic block model (SBM) for its static part with
independent Markov chains for the evolution of the nodes groups through time. We model binary data as well as weighted
dynamic random graphs (with discrete or continuous edges values). Our approach, {motivated by the importance of
controlling  for label switching issues across the different time steps, focuses on detecting
  groups characterized by a stable within group connectivity behavior.} 
We study identifiability of the model parameters, propose an inference procedure based on a variational expectation
maximization algorithm as well as a model selection criterion to select for the number of groups. We carefully discuss
our initialization strategy which plays an important role in the method and compare our procedure with existing ones on 
synthetic datasets. We also illustrate our approach on dynamic contact networks, one 
of encounters among high school students and two others on animal interactions. An implementation
of the method is available as a R package called \texttt{dynsbm}.
 \end{abstract}

\section{Introduction}
Statistical network analysis has become a major field of research, with applications as diverse as sociology, ecology,
biology, internet, etc. General  references on statistical modeling of
random graphs include the recent book by~\cite{Kolaczyk} and the two reviews by~\cite{Goldenberg} and~\cite{Snijders_review}.
While static approaches have been developed as early as in the 60's (mostly in the field of sociology), the literature
concerning dynamic models is much more recent. 
Modeling discrete time dynamic networks is an emerging field that raises many
challenges and we refer to~\cite{Holme_review} for a most recent review. 

An  important part  of the  literature on  static network  analysis is
dedicated to clustering methods, with both
aims of taking into account the intrinsic heterogeneity of the data and summarizing this data through node
classification.  Among  clustering   approaches,  community  detection
methods form a smaller class of methods that aim at
finding groups of highly connected nodes.   Our focus here is not only
on community detection but more generally on node classification based on connectivity behaviors, with a particular interest on model-based
approaches~\citep[see e.g.][]{Matias_Robin_review}.
When considering a sequence of snapshots of a network at different
time steps, these static clustering approaches will give rise to classifications that are difficult to compare  through
time  and thus  difficult to  interpret. An  important thing  to
  note is that  label switching between two successive  time steps may
  not be  solved without an  extra assumption  e.g. that most  of the
  nodes do not change group  across two different time steps. However
  to our knowledge, this kind of assumption has never been discussed in
  the literature. In this work,  we are interested in statistical models for discrete time dynamic
random graphs, with the aim  of providing a node classification varying with 
time, while controlling for label switching  issues across the different time steps. Our answer to this challenge
will be  to focus on the detection of  groups characterized by a stable within group connectivity behavior. We believe
that this is particularly suited for dynamic contact networks. \\

Stochastic block models (SBM) form a widely used class of statistical (and static) random graphs models that provide a 
clustering of the nodes. SBM introduces latent (i.e. unobserved) random variables on the nodes of the graph, taking values in a
finite set. These latent variables represent the nodes groups and interaction between two nodes is governed by these 
corresponding groups. The model includes (but is not restricted to) the specific case of community detection, where
within groups connections have higher probability than across groups ones.
Combining SBM with a Markov structure on the latent part of the process (the nodes classification) is a
natural way of ensuring a smooth evolution of the groups across time and has already been considered in the
literature. 
In~\cite{Yang_etal_ML11}, the authors consider undirected, either binary or finitely valued, discrete time dynamic
random graphs. The static aspect of the data is handled through SBM, while its 
dynamic aspect  is as follows. For each node, its group membership forms a Markov chain, independent of the values of the other nodes memberships. 
There, only the group membership is allowed to vary across time while connectivity parameters among groups
stay constant through time.
The authors propose  a method to infer these parameters
(either online or  offline), based on a combination  of Gibbs sampling
and simulated annealing. 
For binary random graphs,~\cite{Xu_Hero_IEEE} propose to introduce a state-space model through time on (the logit
transform of) the probability of connection between groups. Contrarily
to  the   previous  work,  both  group   membership  and  connectivity
parameters across groups may vary through time. As such, we will see below that this model has a strong
  identifiability problem.  
Their (online) iterative estimation procedure is based on alternating two
steps: a label-switching method to explore the space of node groups configuration, and the (extended) Kalman filter that
optimizes the likelihood when the groups memberships are known. 
Note   that    neither~\cite{Yang_etal_ML11}   nor~\cite{Xu_Hero_IEEE}
propose to infer the number of clusters.
Bayesian variants of these dynamic SB models may be
found for  instance in~\cite{Ishiguro_NIPS2010,herlau13}. 

Surprisingly, we noticed that the above mentioned methods 
were evaluated on synthetic datasets in terms of averaged value over the time steps of a clustering quality index computed
  at fixed time step. Naturally, those indexes do not penalize for label
  switching and two classifications that are identical up to a 
  permutation have the highest quality index value. Computing an 
  index for each time step, the label switching  issue between different time steps disappears and the classification task becomes easier.  Indeed, such criteria
  do not control for  a  smoothed recovery of groups along different time points. It should also
  be noted that the synthetic experiments from these works were performed under the dynamic version of the
  binary affiliation SBM, which has non identifiable parameters. The affiliation SBM,
    also known as planted partition model, corresponds to the case where the connectivity parameter matrix has only two different
    values: a diagonal one that drives within groups connections and an off-diagonal one for across groups connections. 
 In particular, the label  switching issue between different time steps may not be easily removed
  in this particular case. \\
Other approaches for model-based clustering of dynamic random graphs do not rely directly on SBM but rather on variants
of SBM. We mention the random subgraph model (RSM) that combines SBM with the a priori knowledge of a nodes
  partition (inducing subgraphs), by authorizing the groups proportions to differ in the different subgraphs. A dynamic version of RSM that builds upon the approach of~\cite{Xu_Hero_IEEE} appears in~\cite{Zreik_etal}.
Detection of persistent communities has been proposed in~\cite{Liu_persistent} for directed
and dynamic graphs of call counts between individuals. Here the static underlying model is a time and degree-corrected
SBM with  Poisson  distribution  on   the  call
counts. Groups memberships are independent through time instead of Markov, but smoothness in the classification is
obtained by imposing that within groups expected call volumes are constant through time.  Inference is performed through a heuristic greedy
search in the space of groups memberships. Note that only real datasets and no synthetic experiments have been explored in this latter work.

Another very popular statistical method for analyzing static networks is based on latent space models. Each node is
associated to a point in a latent space and probability of connection is higher for nodes whose latent points are
closer~\citep{Hoff_etal}. In~\cite{Sarkar},  a dynamic version of the latent space model is proposed, where the latent
points follow a (continuous state space) Markov chain, with transition kernel given by a Gaussian perturbation on
current position with zero mean and small variance. Latent position inference is performed in two steps: a first initial
guess is obtained through multi dimensional scaling. Then, nonlinear optimization is used to maximize the model
likelihood.  The work by~\cite{Xu_Zheng_09} is very similar, adding a clustering step on the nodes. Finally, \cite{heaukulani13} rely on Monte Carlo Markov
Chain methods to perform a Bayesian inference in a more complicate setup where the latent positions of the nodes are not independent.

Mixed membership models~\citep{Airoldi} are also explored in a dynamic context. The work by~\cite{xing2010}
relies on a state space model for the evolution of the parameters of the priors of both the mixed membership vector of a
node and the connectivity behavior. Inference is carried out through a 
variational Bayes expectation maximisation (\texttt{VBEM}) algorithm~\citep[e.g.][]{Jordan_etal}.

This non exhaustive bibliography on model-based clustering methods for dynamic random graphs shows both the importance
  and the huge interest in the topic.\\

In   the  present   work,  we   explore  statistical   properties  and
frequentist inference  in a model  that combines SBM for  its static
part with independent Markov chains for the evolution of the nodes groups through time. Our approach aims at
  achieving both interpretability and statistical accuracy. 
Our setup is very close to the ones of~\cite{Yang_etal_ML11,Xu_Hero_IEEE}, the first and main difference being that we allow
for  both groups  memberships and connectivity parameters  to vary  through  time. By focusing on groups characterized by a stable within group
connectivity behavior, we are able to ensure parameter identifiability and thus valid statistical inference. 
Indeed, while~\cite{Yang_etal_ML11} use the strong
  constraint of fixed connectivity parameters through time, \cite{Xu_Hero_IEEE} entirely relax this assumption at the (not
  acknowledged) cost of a label switching issue between time steps.
Second, we model binary data as  well as weighted  random graphs, should they be dense or sparse, with discrete
  or continuous edges. Third, we propose a model selection criterion to choose the number of
clusters. To simplify notation, we restrict our
model to undirected random graphs with no self-loops but easy generalizations would handle directed
datasets and/or including self-loops. 

The manuscript is organized as follows. Section~\ref{sec:model} describes the model and sets notation. Section~\ref{sec:ident_heur} gives intuition on the
  identifiability issues raised by authorizing both group memberships and connectivity parameters to freely vary with
  time. This was not pointed out by~\cite{Xu_Hero_IEEE} despite they worked in this context. The section motivates our
  focus on groups characterized by a stable within group connectivity behavior. Section~\ref{sec:ident} then establishes
  our identifiability results. To our knowledge, it is the first dynamic random graph model where
parameters identifiability  (up to  label switching) is  discussed and
established. 
Then, Section~\ref{sec:algo} describes a variational expectation maximization (\texttt{VEM}) procedure for inferring the model 
parameters and clustering the nodes. The \texttt{VEM} procedure works with a fixed number of groups and an Integrated
Classification      Likelihood~\citep[ICL,][]{BCG00} criterion  is proposed   for  estimating the
number  of  groups. 
We  also discuss initialization  of the algorithm   -   an   important   but   rarely   discussed   step,   in
Section~\ref{sec:init}. 
Synthetic         experiments         are         presented         in
Section~\ref{sec:experiments}.   There,   we  discuss   classification
performances  without neglecting  the label  switching issue  that may
occur between time  steps.  
In Section~\ref{sec:real_data}, we illustrate our approach with the analysis of real-life contact networks:  
a dataset of encounters among high school students and two other datasets of animal interactions. We believe that our
model is particularly suited to handle this type of data. 
We mention that the  methods are  implemented into  a R
package available at \url{http://lbbe.univ-lyon1.fr/dynsbm} and will be soon available on the CRAN.
Supplementary Materials (available  at the end of this article)
complete the main manuscript.


\section{Setup and notation}
\subsection{Model description}
\label{sec:model}

We consider weighted  interactions between $N$ individuals recorded through time
in a set of data matrices $\Y=(Y^t)_{1\le t \le T}$.  Here $T$ is the
number of  time points  and for each  value $t\in  \{1,\dots,T\}$, the
adjacency matrix $Y^t=(Y^t_{ij})_{1\le  i\neq j\le N}$ contains real values measuring interactions between
individuals $ i,j\in \{1,\dots,N\}^2$. Without loss of  generality, we consider
undirected  random  graphs without  self-loops,  so  that $Y^t$  is  a
symmetric matrix with no diagonal elements.

We assume that the $N$ individuals are split into $Q$ latent (unobserved)
groups that may vary through time,  as encoded by the random variables
$\Z=(Z^t_i)_{1\le t\le T, 1\le i \le N}$ with values in $\Q^{NT}:=\{1,\dots,Q\}^{NT}$.
This process is modeled as follows. Across individuals, random variables $(Z_i)_{1\le
  i \le N}$ are independent and identically distributed (iid). Now, for  each individual  $i\in
\{1,\dots,N\}$,  the  process  $Z_i=(Z^t_i)_{1\le   t  \le  T}$  is  an irreducible, aperiodic stationary Markov      chain     with      transition      matrix 
$\bpi=(\pi_{qq'})_{1\le q,q'\le Q}$ and initial stationary distribution $\ba=(\alpha_1,\dots,\alpha_Q)$.
When no confusion  occurs, we may alternatively consider  $Z^t_i$ as a
value in $\Q$ or as a random vector $Z^t_i=(Z^t_{i1},\dots,Z^t_{iQ})\in \{0,1\}^Q$
constrained to $\sum_q Z^t_{iq}=1$.

\begin{figure}[h]
  \centering

  \begin{tikzpicture}

 \matrix (m) [matrix of math nodes,row sep=2em, column sep=3em]
    { {}^{\cdots} & Z^{t-1} & Z^{t} & Z^{t+1} & {}^{\cdots} \\
      {}^{\cdots}  & Y^{t-1} & Y^{t} & Y^{t+1} & {}^{\cdots} \\ };

  { [start chain] \chainin (m-1-1);
     {[start branch] \chainin (m-2-1);}
    \chainin (m-1-2);
    {[start branch] \chainin (m-2-2);}
    \chainin (m-1-3);
    { [start branch] \chainin (m-2-3);}
    \chainin (m-1-4);
    { [start branch] \chainin (m-2-4);}
    \chainin (m-1-5);
     {[start branch] \chainin (m-2-5);}
}

  \end{tikzpicture}
\\
\vspace{0.5cm}
    \begin{tikzpicture}
    \matrix (m) [matrix of math nodes, row sep=0.3em, column sep=3em]
    { {}^{\cdots} & Z_1^{t-1} & Z_1^{t} & Z_1^{t+1} & {}^{\cdots}\\
      {}^{\cdots} & Z_2^{t-1} & Z_2^{t} & Z_2^{t+1} & {}^{\cdots}\\
      \vdots & \vdots & \vdots & \vdots & \vdots\\
      {}^{\cdots}& Z_N^{t-1} & Z_N^{t} & Z_N^{t+1} & {}^{\cdots}\\
      &&&&\\
      &&&&\\
      &&&&\\
      &&&&\\
       {}^{\cdots}& Y^{t-1} & Y^{t} & Y^{t+1} &  {}^{\cdots}\\ };

    { [start chain] \chainin (m-1-1);
    \chainin (m-1-2)  [join={node[above,labeled] {\pi}}];
    \chainin (m-1-3) [join={node[above,labeled] {\pi}}];
    \chainin (m-1-4) [join={node[above,labeled] {\pi}}];
    \chainin (m-1-5)  [join={node[above,labeled] {\pi}}];
  } 

    { [start chain] \chainin (m-2-1);
    \chainin (m-2-2) [join={node[above,labeled] {\pi}}];
    \chainin (m-2-3) [join={node[above,labeled] {\pi}}];
    \chainin (m-2-4) [join={node[above,labeled] {\pi}}];
    \chainin (m-2-5) [join={node[above,labeled] {\pi}}];
  }

  { [start chain] \chainin (m-4-1);
    \chainin (m-4-2) [join={node[above,labeled] {\pi}}];
    {[start branch] \chainin (m-9-2)    [join={node[right,labeled] {\phi^{t-1}}}];}
    \chainin (m-4-3) [join={node[above,labeled] {\pi}}];
    {[start branch] \chainin (m-9-3)    [join={node[right,labeled] {\phi^t}}];}
    \chainin (m-4-4) [join={node[above,labeled] {\pi}}];
    {[start branch] \chainin (m-9-4)    [join={node[right,labeled] {\phi^{t+1}}}];}
    \chainin (m-4-5)  [join={node[above,labeled] {\pi}}];
  }

\node [rectangle,draw,minimum height=0.6cm,minimum width=0.8cm] (Yt-) at (m-9-2) {} ;
\node [rectangle,draw,minimum size=0.6cm,] (Yt) at (m-9-3) {} ;
\node [rectangle,draw,minimum size=0.6cm,,minimum width=0.8cm] (Yt+) at (m-9-4) {} ;

\node [
        rectangle,draw,
        above=4mm of Yt-,
	minimum width=0.9cm,
	minimum height=3cm,
] (Zt-) {};

\node [
        rectangle,draw,
        above=4mm of Yt,
	minimum width=0.9cm,
	minimum height=3cm,
] (Zt) {};

\node [
        rectangle,draw,
        above=4mm of Yt+,
	minimum width=0.9cm,
	minimum height=3cm,
] (Zt+) {};

    \end{tikzpicture}
\\
\vspace{0.5cm}
\begin{tikzpicture}

\node (Z1) at (0,1.5) {$Z_1^{t}$};
\node (Z2) at (0.8,1.5) {$Z_2^{t}$};
\node  at (2,1.5){$\cdots$};
\node (Zi) at (2.8,1.5) {$Z_i^{t}$};
\node  at (3.5,1.5){$\cdots$};
\node (Zj) at (4.2,1.5) {$Z_j^{t}$};
\node (D) at (5,1.5){$\cdots$};
\node (ZN-) at (6,1.5) {$Z_{N-1}^{t}$};
\node (ZN) at (7,1.5) {$Z_N^{t}$};
\node (fictif1) at (7,1.3) {};

\node (Y12) at (-0.5,0) {$Y_{12}^{t}$};
\node  at (0.2,0){$\cdots$};
\node (Y1N) at (1,0) {$Y_{1N}^{t}$};
\node (fictif2) at (1,0.3){};
\node  at (2,0){$\cdots$};
\node (Yij) at (3.5,0) {$Y_{ij}^{t}$};
\node  at (4.5,0){$\cdots$};
\node (YN2) at (6,0) {$Y_{N-2,N-1}^{t}$};
\node (YN1) at (7.7,0) {$Y_{N-1,N}^{t}$};

\path[->] 
    (Z1) edge (Y12)
    (Z2) edge (Y12)
    (Z1) edge (Y1N)
    (fictif1) edge (fictif2)
    (Zi) edge (Yij)
    (Zj) edge (Yij)
    (ZN-) edge (YN1)
    (ZN-) edge (YN2)
    (D) edge (YN2)
    (ZN) edge (YN1);

\end{tikzpicture}
  \caption{Dependency  structures  of  the model.   Top:  general  view
    corresponding to hidden Markov model (HMM) structure;
    Middle:  details on  latent structure  organization corresponding
    to $N$ different iid Markov chains $Z_i=(Z_i^t)_{1\le t \le T}$ across individuals; Bottom: details for fixed    time point $t$ corresponding to SBM structure.}
  \label{fig:DAG}
\end{figure}
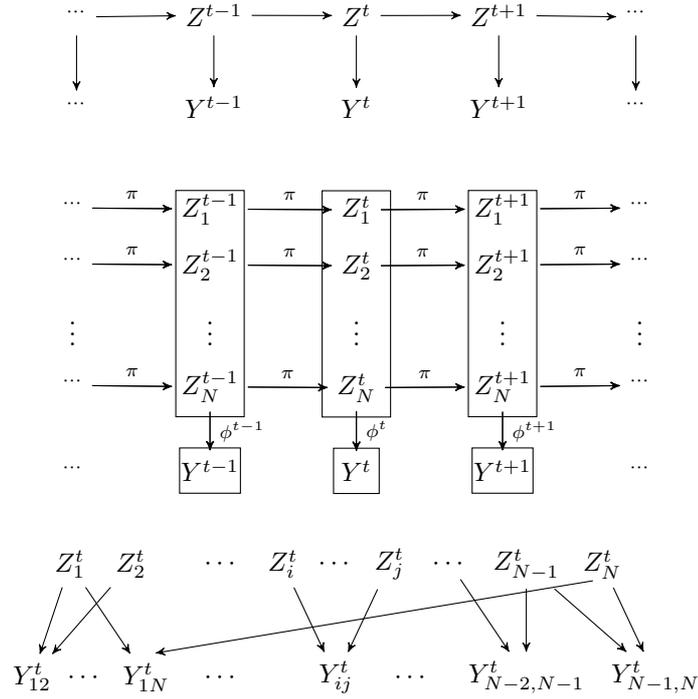

Given latent groups   $\Z$, the  time  varying   random  graphs  $\Y=(Y^t)_{1\le  t   \le
  T}$  are 
independent, the conditional distribution of each $Y^t$ depending only
on $Z^t$.  Then, for fixed $1\le  t\le T$, random graph  $Y^t$ follows a
stochastic block model. In other words, for each time $t$,
conditional on $Z^t$,  random variables $(Y^t_{ij})_{1\le i<j\le N}$
are independent and  the distribution of each  $Y^t_{ij}$ only depends
on $Z^t_i,Z^t_j$.  For now, we  assume a very general  parametric form
for this distribution on $\R$. Following~\cite{Ambroise_Matias}, in order to take into account
  possible sparse weighted graphs, we explicitly introduce a Dirac mass at $0$, denoted by $\delta_0$, as a component of this distribution. 
More precisely, we assume 
\begin{equation}
  \label{eq:emission}
Y^t_{ij}     |     \{Z^t_{iq}Z^t_{jl}=1\}    \sim     (1-\beta^t_{ql})
\delta_0(\cdot) + \beta^t_{ql} F(\cdot,\gamma^t_{ql}), 
\end{equation}
where $\{F(\cdot,\gamma) , \gamma\in \Gamma\}$ is a parametric family of distributions with no point mass at $0$ and
densities (with respect to Lebesgue or counting measure) denoted by $f(\cdot,\gamma)$.  This
could be the Gaussian family with unknown mean and variance, the truncated Poisson family on $\N\setminus\{0\}$ (leading
to a $0$-inflated or $0$-deflated distribution on the edges of the graph), a finite space distribution on $M$
  values (a case which comprises nonparametric approximations of continuous distributions through discretization into a
  finite number of $M$ bins), 
etc. Note that the binary case is encompassed in this setup
with $F(\cdot,\gamma)=\delta_1(\cdot)$, namely the parametric family of laws is reduced to a single point, the Dirac
mass  at 1  and conditional  distribution  of $Y^t_{ij}$  is simply  a
Bernoulli $\mathcal{B}(\beta^t_{ql})$.  In the following and by 
opposition to the 'binary case', we will call 'weighted
case' any setup where the set of distributions $F$ is parametrized and
not reduced to a single point. 
Here, the  sparsity   parameters     $\beta^t=(\beta^t_{ql})_{1\le
  q,l\le     Q}$ satisfy $\beta^t_{ql}\in [0,1]$, with $\beta^t\equiv 1$ corresponding to the particular case of a complete weighted graph.  
As a result of considering undirected graphs, the parameters $\beta^t_{ql},\gamma^t_{ql}$ moreover satisfy
$\beta^t_{ql}=\beta^t_{lq}$                                        and  $\gamma^t_{ql}=\gamma^t_{lq}$ for all $1\le
q,l\leq  Q$.  Note  that for  the moment,  SBM  parameters may  be
different across  time points. We  will go back  to this point  in the
next sections. 
 The model is thus parameterized by 
\[ 
\theta=(\bpi,\bb,\bg)=(\bpi,\{\beta^t,\gamma^t\}_{1\le t\le T})
= (\{\pi_{qq'}\}_{1\le q,q'\le Q},\{\beta^t_{ql},\gamma^t_{ql}\}_{1\le t\le T, 1\le q\le l\le Q}) \in \Theta , 
\]
 and
we let $\pr$ denote the probability distribution on the whole space $\Q^\N \times\R^\N$. 
We also let $\phi(\cdot;\beta,\gamma)$ denote the density of the distribution given
by~\eqref{eq:emission}, namely
\[
\forall y \in \R, \quad \phi(y;\beta,\gamma) = (1-\beta)1\{y=0\} + \beta f(y,\gamma)1\{y\neq 0\},
\]
where $1\{A\}$ is the indicator function of set $A$.
With some abuse of notation and when no confusion occurs, we shorten $\phi(\cdot;\beta^t_{ql},\gamma^t_{ql}) $ to
$\phi^t_{ql}(\cdot)$ or $\phi^t_{ql}(\cdot;\theta)$. 
Directed acyclic  graphs (DAGs) describing  the dependency
structure  of the  variables in  the model  with different  levels of
detail are given in Figure~\ref{fig:DAG}. 
Note that the model assumes that the individuals  are present at any time in the dataset.  An extension that covers for the case  where some nodes are not present at
every  time  point  is given in Section~\ref{sec:extensions} from the Supplementary Materials and used in analyzing the
animal datasets from Section~\ref{sec:animals}.

\subsection{Varying connectivity parameters vs varying group membership}
\label{sec:ident_heur}
\begin{figure}[h]
  \centering
  \begin{minipage}{0.45\linewidth}
   \begin{tikzpicture}
\tikzstyle{every node}=[fill=white]
\tikzstyle{every state}=[text=black,scale=0.8,draw=none]
\node[state] at (1.5,3){${t=t_1}$};

    \tikzstyle{every edge}=[-,>=stealth',shorten >=1pt,auto,thin,draw]
    \tikzstyle{every state}=[text=black,scale=0.4,transform shape]

    \tikzstyle{every node}=[fill=white]
    \node[state] (A1) at (1,1) {\textbf{1}};

    \foreach   \name/\angle/\text   in  {B1/234/\textbf{2},   B2/162/\textbf{3},
        B3/90/\textbf{4}, B4/18/\textbf{5}, B5/-54/\textbf{6}} {
        \node[fill=gray!30!white,state,xshift=9cm,yshift=3.5cm]     (\name)    at
        (\angle:1cm) {\text}; 
      }

      \tikzstyle{every node}=[]  
      \path  (B2) edge (B5)
    (B1) edge (B4);
      \path (B3) edge [bend left]  (B4);
\foreach \from/\to in {1/2,2/3,4/5,5/1}{
     \path (B\from) edge [bend left] (B\to);
      }
    
    \tikzstyle{every node}=[fill=gray!80!white]
    \node[state] (C1) at (0,0) {\textbf{7}};
    \node[state] (C2) at (1,-0.5) {\textbf{8}};
    \node[state] (C3) at (2,0) {\textbf{9}};
    \node[state] (C4) at (-0.5,0.5) {\textbf{10}};
    \node[state] (C5) at (-0.5,1.5) {\textbf{11}};
    \node[state] (C6) at (0,2) {\textbf{12}};
    \path (C1) edge (A1)
    (C2) edge (A1)
    (C3) edge (A1)
    (C4) edge (A1) 
    (C5) edge (A1)
    (C6) edge [bend left] (A1);
    \path (C2) edge [bend right] (C3)
    (C4) edge [bend right] (C6) ; 

    \path (A1) edge [bend right] (B2) 
    (A1) edge [bend left]   (B3) 
    (A1) edge [bend right] (B1);
 
  \end{tikzpicture}
  \end{minipage}
  \begin{minipage}{0.45\linewidth}
   \begin{tikzpicture}
\tikzstyle{every node}=[fill=white]
\tikzstyle{every state}=[text=black,scale=0.8,draw=none]
\node[state] at (1.5,3){$t=t_2$};

    \tikzstyle{every edge}=[-,>=stealth',shorten >=1pt,auto,thin,draw]
    \tikzstyle{every state}=[text=black,scale=0.4,transform shape]

    \tikzstyle{every node}=[fill=white]
    \node[state] (A1) at (1,1) {\textbf{1}};

    \foreach   \name/\angle/\text   in  {B1/234/\textbf{2},   B2/162/\textbf{3},
        B3/90/\textbf{4}, B4/18/\textbf{5}, B5/-54/\textbf{6}} {
        \node[fill=gray!80!white,state,xshift=9cm,yshift=3.5cm]     (\name)    at
        (\angle:1cm) {\text}; 
      }

      \path (B4) edge [bend left] (B5) ;
      \path (A1) edge [bend left] (B3) 
      (A1) edge  (B2) 
      (A1) edge (B4) 
      (A1) edge [bend right]  (B5) 
      (A1) edge [bend right] (B1);

\tikzstyle{every node}=[fill=gray!30!white]
    \node[state] (C1) at (0,0) {\textbf{7}};
    \node[state] (C2) at (1,-0.5) {\textbf{8}};
    \node[state] (C3) at (2,0) {\textbf{9}};
    \node[state] (C4) at (-0.5,0.5) {\textbf{10}};
    \node[state] (C5) at (-0.5,1.5) {\textbf{11}};
    \node[state] (C6) at (0,2) {\textbf{12}};
    \path (C1) edge [bend right] (C2)
     (C2) edge [bend right] (C3)
    (C1) edge [bend left] (C3) 
    (C4) edge [bend right] (C6)
    (C4) edge (C1)
    (C4) edge [bend left] (C5)
    (C1) edge [bend left] (C6)
    (C2) edge (C6)
    (C5) edge (C6); 
    \path (C1) edge (A1)
    (C2) edge (A1)
    (C3) edge (A1)
    (C6) edge [bend left] (A1);

  \end{tikzpicture}
  \end{minipage}
  \caption{Connectivity parameters or group membership variation: a toy example.}
  \label{fig:switch}
\end{figure}
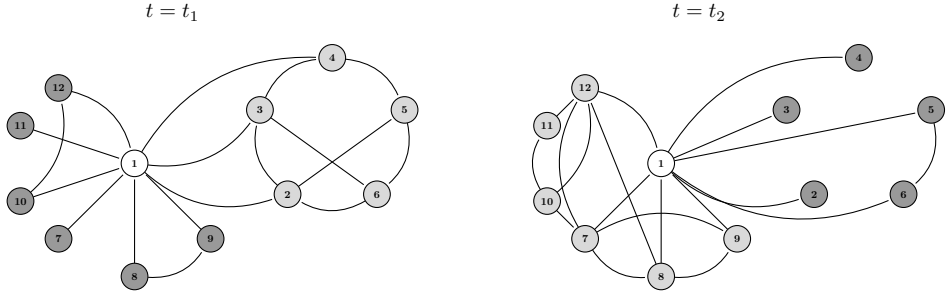

In this section, we give some intuition on why it is not possible to let both connectivity parameters and group
membership vary through time without entering into label switching issues between time steps. To this aim, let us
consider the toy example from Figure~\ref{fig:switch}.

This figure shows a graph between $N=12$ nodes at two different
time points $t_1,t_2$. Node 1 is a hub (namely a highly connected node), nodes 2 to 6 form a community at time $t_1$
(they tend to form a clique) and are peripheral nodes at time $t_2$ and finally nodes 7 to 12 are peripheral at time
$t_1$ while becoming a community at time $t_2$. In observing those two graphs (without the clusters indicated by the
nodes shading), there are at least two possible statistical interpretations relying on a clustering with $Q=3$
groups. The first (illustrated in Figure~\ref{fig:switch}) is to consider that the 3 different groups at stake respectively are: hubs (in white),
a community (light grey) and peripheral nodes (dark grey) and that the nodes 2 to 6 change group from a community to
peripheral group between time $t_1$ and $t_2$ while nodes 7 to 12 change from peripheral group to a community
between those same time points (node 1 stays a hub, in white, for both time points). Another point of view would rather
to consider that nodes 2 to 6 stayed in the same group that was organized as a community at time $t_1$ and is now characterized by
peripheral behavior at time $t_2$, while nodes 7 to 12 also stayed in the same group, behaving peripherally at time
$t_1$ and now as a community at time $t_2$. Obviously none of these two interpretations is better than the
other. Without adding constraints on the model, the label switching phenomenon will randomly output one of these two
interpretations (clustering at time $t_2$ is the same when permuting light grey and dark grey colors). In this context, it is thus 
impossible to recover groups memberships trajectories. We formalize these ideas in the next section through the
concept of parameter identifiability.

The main problem with the previous  example comes from the possibility of
arbitrarily relabeling the groups between two time steps. 
We mentioned in the introduction that a natural idea would be that most of the individuals should not change
  groups between successive time steps. However note that imposing constraints on the transition matrix $\bpi$ (e.g. it
  has large diagonal elements) is useless because estimation would then be unfeasible. Indeed, without imposing zero
  values on the  off diagonal  elements of $\bpi$ (i.e. $Z_i^t$ does not depend on $t$), it can happen that there is no
 labelling of the groups that ensures most individuals stay in the same group. Thus it is not always possible to label
 the groups so  that between two successive time  steps, estimation of
 the transition parameters would be constrained to have
 large diagonal elements. Thus we choose to focus our attention
 on groups characterized through their stable within group connectivity parameter. This choice is reminiscent of works
 on detection of persistent communities~\citep{Liu_persistent}, except that we do not restrict our attention to
 communities  (i.e. groups  of highly  connected individuals).  In the
 above  toy  example, this  corresponds  to  the first  interpretation
 rather than the second. 
Other choices could be made and we believe that this one is particularly suited to model social networks or contact data where the groups are
 defined as  structures exhibiting a stable  within group connectivity
 behavior and individuals may change groups through time (see Section~\ref{sec:real_data} for applications on real datasets).


\subsection{Parameters identifiability}
\label{sec:ident}
Let   us  recall   that   with  discrete   latent  random   variables,
identifiability can only be obtained up to a label switching on
the node groups $\Q$. For any permutation $\sigma$ in $\mathfrak{S}_Q$ (the set of permutations on
$\Q$) and any $\theta \in \Theta$, we define 
\[
\sigma(\theta):= (\{\pi_{\sigma(q)\sigma(q')}\}_{1\le q,q'\le
  Q},\{\beta^t_{\sigma(q)\sigma(l)},\gamma^t_{\sigma(q)\sigma(l)}\}_{1\le t\le T, 1\le q\le l\le Q}) . 
\] 
It  should  be   noted  that  here,  the   permutation  $\sigma$  acts
\emph{globally}, meaning that it is the same at each time point $t$. 
Now, if we let $\pr^Y$ denote the marginal of $\pr$ on the set of observations
$\Y$,   
identifiability of the parameterization,  up to label switching means 
\[
\forall \theta, \tilde \theta \in \Theta, \quad \pr^Y= \mathbb{P}_{\tilde \theta}^Y  \implies \exists \sigma \in \mathfrak{S}_Q, \theta =
\sigma(\tilde \theta).
\]
Without additional constraints on  the transition matrix $\bpi$ or
on the parameters $(\bb,\bg)$, the  parameters may not be recovered up
to label switching. However, it could be that the static SBM part of the parameter is 
recovered up to a \emph{local} label switching. Local label
switching on SBM part of the parameter is the weaker following property 
\[
\forall   \theta,    \tilde   \theta   \in   \Theta,    \quad   \pr^Y=
\mathbb{P}_{\tilde \theta}^Y \implies \exists \sigma_1,\dots, \sigma_T
\in \mathfrak{S}_Q^T, \forall t, (\beta^t,\gamma^t) =
\sigma_t(\tilde \beta^t,\tilde \gamma^t).
\]
This property is not satisfactory since clustering in models that only
satisfy a local identifiability of SBM part of the parameter prevents from
obtaining a  picture of the  evolution of  the groups across  time.

A  formal example of the fact  that if both $Z^t$ and 
$(\beta^t,\gamma^t)$ may vary through time, then the parameter can not
be identified up to  label switching without additional constraints is given in Section~\ref{ex:nonident} from Supplementary Materials.
We stress that this example implies that dynamic affiliation SBM (or planted partition model)   does   not    have
identifiable parameters and groups may not be recovered consistently across time. This
is an important point as previous authors have tried to recover groups  from this type of synthetic  datasets and
evaluated their estimated classification in a non natural way.

As a consequence and following the ideas developed in Section~\ref{sec:ident_heur}, we choose to impose the following constraints  on the parameter $\theta$ 
\begin{equation}
  \label{eq:intra_gp_const}
\forall q \in \Q, \forall t,t'\in \{1,\dots, T\}, \quad 
\left\{
  \begin{array}{ll}
 \text{Binary case:} & \beta^{t}_{qq}= \beta^{t'}_{qq} :=\beta_{qq},\\
\text{Weighted case:} & \gamma^{t}_{qq}= \gamma^{t'}_{qq} :=\gamma_{qq}. 
  \end{array}
\right .
\end{equation}
Under the above condition, we  focus on groups characterized by a stable within group connectivity
  behavior ($\beta_{qq}$ or $\gamma_{qq}$ is constant with time). 
Note that this constraint could in fact be weakened as follows 
\begin{equation}
  \label{eq:gp_const}
\forall q \in \Q, \forall t,t'\in \{1,\dots, T\}, \exists l \in  \{1,\dots, T\},\quad 
\left\{
  \begin{array}{ll}
 \text{Binary case:} & \beta^{t}_{ql}= \beta^{t'}_{ql} ,\\
\text{Weighted case:} & \gamma^{t}_{ql}= \gamma^{t'}_{ql}. 
  \end{array}
\right .
\end{equation}
In this latter condition, the group $l$  that helps characterizing the group $q$ between the two different time points $t,t'$
may depend on $q,t,t'$. Such a constraint may be useful if groups are not characterized by a stable within group
connectivity but rather by their connectivity to at least one specific other
group. Note that  for estimation purposes, this group $l$  needs to be
known in advance (for each $q,t,t'$) which requires a more complex
a priori modeling of the data. In the following, we choose to restrict our
attention to constraint~\eqref{eq:intra_gp_const} only but our theoretical results remain valid under~\eqref{eq:gp_const}.    
We prove below that these constraints, combined with the same conditions used for identifiability in the static
  case, are sufficient to ensure identifiability of the parametrization in our dynamic setup.

\begin{hyp}[Weighted case]
\label{hyp:ident}
  We assume that 
  \begin{itemize}
  \item[i)] For any $t\ge 1$, the $Q(Q+1)/2$ values $\{\gamma^t_{ql}, 1\le q\le l\le Q\}$ are distinct, 
\item[ii)] The family of distributions $\mathcal{F}=\{f(\cdot,\gamma), \gamma \in \Gamma\}$ is such that all elements $f(\cdot, \gamma)$
  have no point mass at $0$ and the parameters of finite mixtures of distributions in $\mathcal{F}$ are identifiable, up
  to label switching. 
  \end{itemize}
\end{hyp}
Assumption~\ref{hyp:ident} is the condition that ensures identifiability of static weighted SBM~\cite[see Theorem
  12 in][]{AMR_JSPI}. 
Note that it does not impose any constraint on the sparsity parameters $\beta_{ql}^t$ in the weighted case. In
particular and for parsimony reasons, these may be chosen identical (to some $\beta^t$ or some constant $\beta$) or set
to two different values, e.g. $\beta^t_{qq}=\beta^t_{\text{in}}$ and $\beta^t_{ql}=\beta^t_{\text{out}}$ whenever $q\neq
l$ at each time point (or even constant with time).

\begin{prop}\label{prop:ident}
Considering the distribution $\pr^Y$ on the set of observations and assuming the constraint~\eqref{eq:intra_gp_const}, the parameter $\theta=(\bpi,\bb,\bg)$ satisfies the following: 
\begin{itemize}
\item \textbf{Binary case:}  $\theta$ is generically identified from $\pr^Y$, up to label switching, as soon as $N$ is not too small with respect to
  $Q$, 
\item \textbf{Weighted case:} Under additional Assumption~\ref{hyp:ident}, the parameter $\theta$ is identified from $\pr^Y$, up to
  label switching, as soon as $N\ge 3$.
\end{itemize}
\end{prop}
Generic identifiability means 'up to excluding a subset of zero Lebesgue measure of the parameter set'. 
We refer to~\cite{AMR_AoS,AMR_JSPI} for more details.
In particular for the binary case, assuming that the matrix of Bernoulli parameters $\bb$ has distinct rows is a generic
constraint (meaning that it removes a subset of zero Lebesgue measure of the parameter set). As we do not specify the whole generic constraint that
is needed here, we do not stress that one either. But the reader should have it in mind in the binary setup. 
Finally, note that the condition on the number of nodes $N$ being not too small
in   the   binary  case   is   given   precisely  in Theorem   2
from~\cite{AMR_JSPI}.  The  particular  affiliation  case (planted partition)  is  not
covered by  these results and further  discussed in Section~\ref{sec:affil_ident}
from Supplementary Materials.

\begin{proof}
  The  proof  combines  the approaches  of~\cite{leroux}  for  proving
  identifiability of hidden Markov models (HMM) parameters
  and~\cite{AMR_JSPI} that studies identifiability for (static) SBM.

First, we fix a time point $t\ge 1$ and consider the marginal distribution $\pr(Y^t)$. According to Theorems 1,2 (binary
case with $Q=2$ and $Q\ge 3$, respectively) and Theorem 12 (weighted case) in~\cite{AMR_JSPI} on parameters identifiability in static
SBM, there exists a permutation $\sigma_t$ on the group labels $\Q$ such that we can identify 
$(\beta^t,\gamma^t)$ as well as the marginal distribution $\ba$, up to this permutation. This result stands
generically in the binary case only. 

Now, for two different time points $t,t'$, we use the
constraint~\eqref{eq:intra_gp_const} and the assumption of distinct parameter values in order to identify the parameters
$\{(\beta^t,\gamma^t) , t \ge 1\}$ up to a (common) permutation $\sigma$ on $\Q$. 
Indeed, in the binary case, assuming that the within groups Bernoulli parameters satisfy $\beta^t_{qq}=\beta^{t'}_{qq}$ and that
the set $\{\beta^t_{qq} ; 1\le q\le Q\}$ contains $Q$ distinct values (a generic constraint) suffices to obtain a global permutation
$\sigma$, not depending on time $t$, up to which $\{(\beta^t,\gamma^t), t\ge 1\}$ are identified. The same applies in
the weighted case, by assuming equality between the parameter  $\gamma_{qq}^t =\gamma_{qq}^{t'}$ for any $t,t'$.

It  remains to identify the  transition matrix $\bpi$ (up  to the
same permutation $\sigma$). We fix an edge $(i,j)$ and following~\cite{leroux}, consider the bivariate
distribution $\pr(Y^t_{ij},Y^{t+1}_{ij})$. This is given by 
\begin{equation}
  \label{eq:joint}
\pr(Y^t_{ij},Y^{t+1}_{ij}) =\sum_{q_1,q_2,l_1,l_2 \in \Q} \alpha_{q_1}\alpha_{l_1} \pi_{q_1q_2} \pi_{l_1l_2}
\phi^t_{q_1l_1}(Y^t_{ij})\phi^{t+1}_{q_2l_2}(Y^{t+1}_{ij}). 
\end{equation}
Note that~\cite{Teicher67} has proved the equivalence between parameters identifiability of the mixtures of a family of
distributions and parameters identifiability of the mixtures of finite
products from  this same family. For  the sake of clarity,  we develop
his proof adapted to our context. We thus write 
\[
\pr(Y^t_{ij},Y^{t+1}_{ij}) =\sum_{q_2,l_2\in \Q} \Big(\sum_{q_1,l_1 \in \Q} \alpha_{q_1}\alpha_{l_1} \pi_{q_1q_2} \pi_{l_1l_2}
\phi^t_{q_1l_1}(Y^t_{ij}) \Big) \phi^{t+1}_{q_2l_2}(Y^{t+1}_{ij}). 
\]
As the  mixtures from the  family $\{\phi^{t+1}_{ql}, 1\le q\le  l \le
Q\}$ have identifiable parameters (Assumption~\ref{hyp:ident}, ii)), we can
identify the mixing distribution 
\[
\sum_{q_2,l_2\in \Q} \Big(\sum_{q_1,l_1 \in \Q} \alpha_{q_1}\alpha_{l_1} \pi_{q_1q_2} \pi_{l_1l_2}
\phi^t_{q_1l_1}(Y^t_{ij})                                        \Big)
\delta_{(\beta^{t+1}_{q_2l_2},\gamma^{t+1}_{q_2 l_2})} (\cdot). 
\]
Now, applying again this identifiability at time $t$ and constraint~\eqref{hyp:ident},  we may  identify the  whole  mixing
distribution 
\[
\sum_{q_2,l_2\in  \Q} \sum_{q_1,l_1  \in \Q}  \alpha_{q_1}\alpha_{l_1}
\pi_{q_1q_2} \pi_{l_1l_2}
\delta_{(\beta^t_{q_1 l_1},\gamma^t_{q_1 l_1})} (\cdot)\otimes 
\delta_{(\beta^{t+1}_{q_2l_2},\gamma^{t+1}_{q_2,l_2})} (\cdot). 
\]
This proves that the mixture given by~\eqref{eq:joint} has identifiable components. 
From  this   mixture and the fact that we already identified the parameters $(\bb,\bg)$  up to a global permutation, we may extract the set of coefficients $\{\alpha_q^2\pi_{qq'}^2 , 1\le q,q'\le Q\}$ that corresponds
to        the         components        $\phi^t_{qq}\phi^{t+1}_{q'q'}$
in~\eqref{eq:joint}. As we also already obtained the values $\{\alpha_q,
1\le q \le Q\} $, this now identifies the parameters $\{\pi_{qq'} , 1 \le q,q'\le Q\}$.   This concludes the proof. 
\end{proof}



\section{Inference algorithm}
\label{sec:algo}
\subsection{General description}
As usual with latent variables, the log-likelihood $\log\pr(\Y)$ contains a sum over all possible latent configurations
$\Z$ and thus may  not be computed except for small  values of $N$ and
$T$. A classical solution is to rely on
expectation-maximisation (\texttt{EM}) algorithm~\citep{DempsterLR}, an iterative procedure that finds local maxima of the log-likelihood. The use of
\texttt{EM} algorithm relies on the computation of the conditional distribution of the latent variables $\Z$ given the
observed ones $\Y$.  However in the context  of stochastic block model,
this distribution has not a factored form and thus may
not be computed efficiently. A classical solution is to rely on variational approximations of \texttt{EM} algorithm~\citep[\texttt{VEM}, see for instance][]{Jordan_etal}. These approximations have been first proposed in the context of
SBM in  \cite{Daudin_etal08} and  later developed in  many directions,
such as  online procedures~\citep{Zanghi_PatternRec08,Zanghi_AAS10} or
Bayesian  \texttt{VEM}~\citep{Latouche_Bayes_EM}.   We  refer  to  the
review by~\cite{Matias_Robin_review} for 
more   details  about   \texttt{VEM}   algorithm   (in  particular   a
presentation  of   \texttt{EM}  viewed   as  a  special   instance  of
\texttt{VEM})  and its  comparison to  other estimation  procedures in
SBM. Note  that convergence properties of  \texttt{VEM} algorithms are
discussed  in full  generality in~\cite{cv_varEM}  and in  the special
case of SBM in~\cite{Celisse_etal,Bickel_etal}.

\paragraph*{\texttt{VEM} for dynamic SBM.}
In our context of dynamic random graphs, we start by writing the complete data log-likelihood of the model
\begin{align}
  \label{eq:complete_log_lik}
  \log \pr(\Y,\Z) =&  \sum_{i=1}^N \sum_{q=1}^Q Z^1_{iq} \log \alpha_q 
 + \sum_{t=2}^{T}  \sum_{i=1}^N \sum_{1\le q,q'\le Q} Z^{t-1}_{iq}Z^{t}_{iq'} \log \pi_{qq'} \nonumber\\
& + \sum_{t=1}^{T}  \sum_{1\le i <j\le N} \sum_{1\le q,l\le Q} Z^t_{iq}Z^{t}_{jl} \log \phi(Y^t_{ij}; \beta^t_{ql},\gamma^t_{ql}) .
\end{align}
We now explore the dependency structure of the  conditional  distribution  $\pr(\Z|\Y)$.  First, note  that it
can be easily deduced  from the DAG of the model (Figure~\ref{fig:DAG}, top) that 
\[
\pr(\Z|\Y) = \pr(Z^1|Y^1) \prod_{t=2}^T\pr(Z^t|Z^{t-1}, Y^t). 
\]
However, the distribution  $\pr(Z^t|Z^{t-1}, Y^t)= \pr((Z^t_{i})_{1\le i\le N}|Z^{t-1}, Y^t) $ can not be further
factored. Indeed, for any $i\neq j$, the variables $Z^t_i,Z^t_j$ are not independent when conditioned on $Y^t$. Our
variational approximation  naturally considers the following  class of
probability distributions $\qr:=\qr_{\tau}$ parameterized by $\tau$ 
\begin{align*}
\qr_\tau(\Z) & =   \prod_{i=1}^N \qr_\tau (Z_i) = \prod_{i=1}^N  \qr_\tau(Z^1_i)\prod_{t=2}^T\qr_\tau
(Z^t_i|Z^{t-1}_i)  \\
&=         \prod_{i=1}^N     \Big[\prod_{q=1}^Q
\tau(i,q)^{Z^1_{iq}} \Big] \times  \prod_{t=2}^T \prod_{1\le q,q'\le
  Q} \tau(t,i,q,q')^{Z^{t-1}_{iq}Z^{t}_{iq'}} ,
\end{align*}
where  for   any  values  $(t,i,q,q')$,  we   have  $\tau(i,q)$  and
$\tau(t,i,q,q')$ both belong to the set $[0,1]$ and are constrained by 
$\sum_{q} \tau(i,q)=1$ and  $\sum_{q'}\tau(t,i,q,q')=1$.  This class
of probability distributions $\qr_{\tau}$
corresponds to considering independent laws through individuals, 
while for each $i\in \{1,\dots,N\}$, the distribution of $Z_i$ under $\qr_\tau$ is
the one  of a  Markov chain (through  time $t$),  with inhomogeneous
transition    $\tau(t,i,q,q')=\qr_{\tau}(Z^t_{i}=q' |Z^{t-1}_{i}=q)$     and    initial    distribution
$\tau(i,q)=\qr_{\tau}(Z^1_{i}=q)$. 

We will need  the marginal components  of $\qr_{\tau}$, namely
$\taum(t,i,q) :  =\qr_{\tau}(Z^t_i=q)$. These quantities  are computed
recursively by 
\[
\taum(1,i,q) = \tau(i,q) \text{ and } \forall t\ge 2, \, \taum(t,i,q) =
\sum_{q'=1}^Q \taum(t-1,i,q')\tau(t,i,q',q).
\]
Note  also  that  all  these   values  $\taum(t,i,q)$  depend  on  the
initial distribution $\tau(i,q)$.  Entropy of $\qr_\tau$ is
denoted by $\h(\qr_\tau) $.
Using this class of probability distributions on $\Q^\N$, \texttt{VEM}
algorithm  is  an  iterative   procedure  to  optimize  the  following
criterion 
\begin{align}
  \label{eq:J}
  J(\theta,\tau) := & \esp_{\qr_\tau}(\log \pr(\Y,\Z)) + \h(\qr_\tau) \nonumber\\
=&  \sum_{i=1}^N \sum_{q=1}^Q \tau(i,q) [\log \alpha_q -\log \tau(i,q)]\nonumber\\ 
& + \sum_{t=2}^{T} \sum_{i=1}^N  \sum_{1\le q,q'\le Q} \taum(t-1,i,q)
  \tau(t,i,q,q') [\log \pi_{qq'} -\log \tau(t,i,q,q')] \nonumber\\
& + \sum_{t=1}^{T}  \sum_{1\le i<j\le N} \sum_{1\le q,l\le Q} \taum(t,i,q) \taum(t,j,l) \log \phi^t_{ql}(Y^t_{ij}) . 
\end{align}
It consists in iterating the following two steps. At $k$-th iteration, with current parameter
value $(\tau^{(k)},\theta^{(k)})$, we do 
\begin{itemize}
\item \texttt{VE}-step: Compute $\tau^{(k+1)}=\argmax_{\tau} J(\theta^{(k)},\tau)$,
\item \texttt{M}-step: Compute $\theta^{(k+1)}=\argmax_{\theta} J(\theta,\tau^{(k+1)})$.
\end{itemize}

\begin{prop}
\label{prop:VEM}
  The  value  $\hat \tau  $  that  maximizes  in $\tau$  the  function
  $J(\theta,\tau)$ satisfies the fixed point equation 
\begin{align*}
 \forall t\ge 2, \forall i\ge 1,\forall q,q'\in \Q, \quad \hat \tau(t,i,q,q')\propto \pi_{qq'} \prod_{j=1}^N \prod_{l'=1}^Q
 [\phi^t_{q'l'}(Y^t_{ij})]^{\htaum(t,j,l') },
\end{align*}
where $\propto$ means 'proportional to' (the constants are obtained by
the constraints on $\tau$). 
Moreover, the value $(\hat \bpi,\hat\bb)$ that  maximizes  in $(\bpi,\bb)$  the  function
  $J(\theta,\tau)$ satisfies
\begin{align*}
  \forall (q,q')\in \Q^2,\quad \hat \pi_{qq'} &\propto  \sum_{t=2}^T
 \sum_{i=1}^N \taum(t-1,i,q) \tau(t,i,q,q') ,\\
 \forall t, \forall q\neq l \in \Q^2,\quad \hat \beta^t_{ql} &= \frac {\sum_{i,j}
  \taum (t,i,q) \taum(t,j,l) 1_{Y^t_{ij}\neq 0} } { \sum_{i,j}\taum(t,i,q) \taum(t,j,l)}, \\
\forall q \in \Q,\quad \hat \beta_{qq} & = \frac {\sum_t\sum_{i,j}
  \taum(t,i,q) \taum (t,j,q) 1_{Y^t_{ij} \neq 0} } { \sum_{t,i,j}\taum (t,i,q) \taum (t,j,q) } . 
\end{align*}
\end{prop}
The proof  of this  result is immediate  and omitted.  
Note that we have given a formula with constant (through time) values
  $\beta_{qq}$ for  any group $q\in  \Q$. While this assumption  is an
  identifiability requirement in the binary setup, it is not necessary
  in the weighted
  case. In this latter case, we use it only for parsimony reasons. The corresponding formula
  when this parameter is not assumed to be constant may be easily
  obtained. 

To complete the algorithm's description, we provide equations to
update the parameters $\tau(i,q)$ and $\alpha_q$ of initial distributions as well as the connectivity parameter $\bg$. 
First, optimization of $J(\theta,\tau)$   with  respect   to  the   initialization  parameters
$\tau(i,q)$ is a little bit more involved.  By neglecting the dependency on $\tau(i,q)$ of some
terms appearing  in criterion $J$, we  choose to update this  value by
solving the fixed point equation 
\begin{align}
\label{eq:approx}
  \forall i\ge 1, \forall q\in \Q, \quad \hat \tau(i,q) &\propto\alpha_q \prod_{j=1}^N \prod_{l=1}^Q
  \phi^1_{ql}(Y^1_{ij})^{\hat \tau(j,l)} .
\end{align}
Our experiments show that this is a reasonable approximation (Section~\ref{sec:experiments}). For the sake of
completeness, we provide in Section~\ref{sec:annexe} from Supplementary Materials the exact equation satisfied by the solution. 

Now parameter $\ba$ is not obtained from maximizing $J$ as it
is not a free parameter but rather the stationary distribution associated
with transition  $\bpi$. Thus, $\ba$  is obtained from  the empirical mean of the marginal
distribution $\htaum$ over all data points 
\begin{equation*}
  \forall   q  \in   \Q,  \quad   \hat  \alpha_q     =  \frac   1  {NT}
   \sum_{t=1}^T\sum_{i=1}^N  \htaum(t,i,q) . 
\end{equation*}
Finally, optimization with respect to $\bg$ depends on the choice of the parametric family $\{f(\cdot,\gamma), \gamma \in
\Gamma\}$.      
We  provide  explicit  formulae for the most widely used families of conditional distributions on the edges (binary or weighted case)
 in Section~\ref{sec:examplesM} from Supplementary Materials. More precisely, we give these formulae for Bernoulli,
 finite space, (zero-inflated or deflated) Poisson and Gaussian homoscedastic distributions.

\begin{remark}
Performing \texttt{EM} algorithm in HMM (Figure~\ref{fig:DAG}, top) requires the use of forward-backward equations in
order to deal with transition terms $Z^{t-1}_{iq}Z^t_{iq}$ appearing in the complete data
log-likelihood~\eqref{eq:complete_log_lik}. In our setup, forward-backward equations are useless and replaced by a
variational approximation. Indeed, it can be seen from Figure~\ref{fig:DAG}, middle, that the conditional distribution
of $Z^{t-1}_{iq}Z^t_{iq}$ given the data can not be computed exactly through such forward-backward equations. This is
due to the fact that the variables $Y^t=\{Y_{ij}^t\}_{i,j}$ depend on all hidden ones $Z^t_1,\dots,Z^t_N$ and focusing only on
$Z^t_i$ is not sufficient to determine their distribution.  
\end{remark}

\begin{remark}
In~\cite{Yang_etal_ML11}, the authors  derive a \texttt{VEM} procedure
in a similar (slightly less general) setup, but their variational approximation uses independent
marginals  (through   individuals  and   also  time  points).    As  a
consequence, the \texttt{VE}-step that they derive is more involved
than ours \citep[see Section 4 in][]{Yang_etal_ML11}. 
\end{remark}

\paragraph*{Model selection.}  
Model selection on  the number of groups $Q$ is  an important step. In
case of latent variables, when the true data likelihood may not be easily
computed, model selection may be done by maximizing an integrated
classification           likelihood          (ICL)           criterion
\citep{BCG00}. For any number of  groups $Q\ge 1$, let $\hat \theta_Q$
be the estimated  parameter value with $Q$ groups and  $\hat \Z$ the
corresponding maximum a posteriori (MAP) classification at $\hat \theta_Q$. 
In our case, the general form of ICL is given by
\begin{equation}
  \label{eq:ICL_gen}
ICL(Q) = \log \mathbb{P}_{\hat \theta_Q}(\Y,\hat \Z) -\frac 1 2 Q(Q-1)\log[ N(T-1)] -pen(N,T,\bb,\bg),
\end{equation}
where   the first  penalization term accounts for transition
matrix $\bpi$ and $pen(N,T,\bb,\bg)$ is a penalizing term for the connectivity parameters
$(\bb,\bg)$. As the number of parameters $(\bb,\bg)$ depends on the specific form of the family $\{f(\cdot;\gamma),
\gamma \in \Gamma\}$, we provide context dependent expressions for ICL
in Section~\ref{sec:examplesM} from Supplementary Materials (along with the expressions of parameter estimates from the
\texttt{M}-step for each case considered). 
Note  that the first penalization  term accounts
for $N(T-1)$ latent transitions while the number of observations corresponding
to SBM part of the parameter in $pen(N,T,\bb,\bg)$ will be different.  We refer
to~\cite{Daudin_etal08} for  an expression of  ICL in static  SBM that
shows  an analogous  difference  in penalizing  groups proportions  or
connectivity parameters. 
Note that there are no theoretical results on the convergence of the ICL procedure (neither in simple mixture models nor in the SBM
case). However the criterion shows very good performances on synthetic experiments and is widely used (see
Section~\ref{sec:experiments} for experiments in our setup). Nonetheless we mention that the criterion is not suited in
the case of a finite space conditional distribution (see Example~\ref{ex:finite} in Section~\ref{sec:examplesM} from
Supplementary Materials for more details).

\subsection{Algorithm initialization}
\label{sec:init}
All  \texttt{EM}  based procedures  look  for  local maxima  of  their
objective  function  and careful  initialization  is  a key  in  their
success.  
For static SBM, \texttt{VEM} procedures often rely on a
\texttt{k-means}  algorithm  on  the  adjacency matrix  to  obtain  an
initial clustering of the individuals. In our context, the dynamic aspect of the data needs
to  be properly  handled.  We choose  to  initialize our  \texttt{VEM}
procedure by  running \texttt{k-means} on  the rows of  a concatenated
data  matrix containing  all the  adjacency time  step matrices  $Y^t$
stacked  in  consecutive  column  blocks. As  a  result,  our  initial
clustering of the individuals is  constant across time (namely $Z_i^t$
does not  depend on $t$).  A consequence of  this choice is  that this
initialization works well when the  groups memberships do not vary too
much across  time (see Section~\ref{sec:experiments} where we explore different values of transition matrix $\bpi$).  
In practice, real-life contact networks will either exhibit nodes that do not change group at all (see Section~\ref{sec:real_data})
or nodes that leave a group and then come back to this group. Our initialization is performant in these cases. 
Another consequence is that while we would expect
the performances of the procedure to increase with the number
$T$ of  time steps, we sometimes observe on the contrary a decrease in these performances. 
This is due to the fact that
increasing $T$  also increases  the probability  for an  individual to
change group at  some point in time and thus  starting with a constant
in time  clustering of the  individuals, it becomes more  difficult to
correctly    infer    the    groups   membership    at    each    time
point  (see in  Section~\ref{sec:experiments}  the difference  between
results  for $T=5$  and $T=10$).

To conclude this section, we mention that initialization is also a crucial point for other methods and we discuss
  in the next section its impact on the algorithm proposed in~\cite{Yang_etal_ML11}.


\section{Synthetic experiments}
\label{sec:experiments} 

The  methods presented  in this  manuscript are  implemented into  a R
package and  available at  \url{http://lbbe.univ-lyon1.fr/dynsbm}.
While the complexity of the estimation algorithm is $O(T  Q^2  N^2)$,
  the computation time remains acceptable for networks with a few thousands of nodes (see Supplementary Figure~\ref{fig:cputime}).

\subsection{Clustering performances}
In  this  section, we  explore  the  performances  of our  method  for
clustering the nodes across the different  time steps. To this aim, we
will consider  two different  criteria. We rely  on the  adjusted Rand
index~\citep[ARI][]{HA1985}  to  evaluate  the agreement  between  the
estimated and the true latent structure. This index is smaller than 1, two identical  latent structures (up to label  switching) having an
ARI equal to 1. Note that it can take negative values and is built on Rand index with a correction for chance.
 Now there are two different ways of
using      ARI      in       a      dynamic      setup.      Following
\cite{Yang_etal_ML11,Xu_Hero_IEEE},  we  first  consider  an  averaged
value over the different time steps  $1\le t\le T$ of ARI$_t$ computed
at time $t$. In this approach the dynamic setup may be viewed as a way
of improving the node clustering at  each time step over a method that
would cluster  separately the nodes  at each time step.  However, this
averaged index does not say anything about the smooth recovery of group
memberships along  time. In  particular, it  is invariant  under local
switching     on     SBM     part    of     the     parameter     (see
Section~\ref{sec:ident}). Thus  we also consider the  global ARI value
that compares the  clustering of the set of nodes  for all time points
with the true latent structure.  Obviously, good performances for this
criteria are more difficult to obtain.

We use  synthetic datasets  created as  follows. We
consider binary graphs with $N=100$ nodes and $T\in \{5;10\}$
different  time  steps.  We  assume $Q=2$  latent  groups  with  three
different values for the transition matrix $\bpi$
\begin{equation*}
\bpi_{low}=  \begin{pmatrix}
    0.6 & 0.4 \\
0.4 & 0.6 
  \end{pmatrix} ; 
\bpi_{medium}=    \begin{pmatrix}
    0.75 & 0.25 \\
0.25 & 0.75 
  \end{pmatrix} ; 
\bpi_{high}=    \begin{pmatrix}
    0.9 & 0.1 \\
0.1 & 0.9 
  \end{pmatrix} .
\end{equation*}

These three cases correspond respectively to \emph{low, medium} and \emph{high}
group stability. Namely in the first case, individuals are more likely
to change group across time, resulting in a more difficult problem from the point of view of the initialization of our
algorithm (see Section~\ref{sec:init}).  The stationary distribution in
those three  cases is  $\ba=(1/2,1/2)$ so that  the two  groups have
similar proportions.

 \begin{table}
   \caption{Bernoulli parameter values in 4 different cases, plus an affiliation example.}
  \centering
  \begin{tabular}{|c|c|c|c|}
    \hline 
     Easiness  & $\beta_{11}$ & $\beta_{12}$& $\beta_{22}$\\
    \hline 
    low-&0.2&0.1& 0.15\\
    low+&0.25&0.1& 0.2\\
    medium-&0.3&0.1&0.2 \\
    medium+&0.4&0.1&0.2 \\
    \hline 
    med w/ affiliation&0.3&0.1& 0.3\\
    \hline 
  \end{tabular}
   \label{tab:param} 
 \end{table}

As for the Bernoulli parameters $\bb$, we  explore 4 different cases (see Table~\ref{tab:param}) representing
different difficulty levels, plus a specific example of affiliation 
for which  we recall  that parameters are  not identifiable in the dynamic setting. We note that this latter case
  satisfies the separability condition   established for sparse planted partition models \citep[see][for more details]{Mossel2014}.
This means that static reconstruction of the groups is conjectured to be possible
  (and we consider that this static problem corresponds to a medium difficulty).

\begin{figure}
 \centering
  \includegraphics[width=\textwidth,height=6cm]{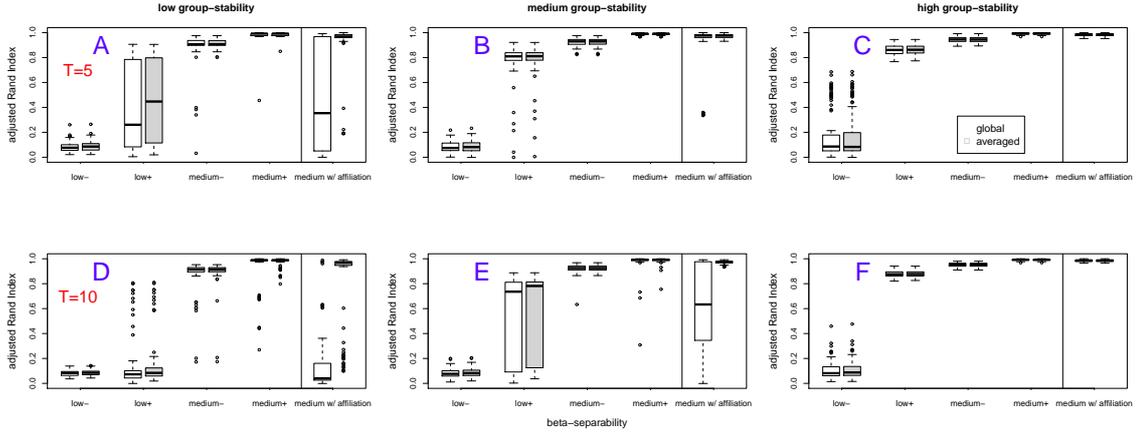}
  \caption{Boxplots  of global  ARI  (white, left)  and averaged  ARI
    (grey, right) in different setups. From left to right: the
    three panels  correspond to $\bpi =  \bpi_{low}$ (panels A,D), $\bpi_{medium}$ (panels B,E) and
    $\bpi_{high}$ (panels C,F), respectively.  In each  panel, from left  to right:
    results  corresponding  to  $\bb=low-,low+,medium-,medium+$
    and affiliation case, respectively. First row: $T=5$ time points, second row: $T=10$.}
 \label{fig:ARI} 
 \end{figure}

For  each  combination  of  $(\bpi,\bb)$, we  generate  100  datasets,
estimate  their parameters,  cluster  their nodes  and  report in Figure~\ref{fig:ARI}  boxplots of  a
global and of an averaged ARI value.

Figure~\ref{fig:ARI} confirms that it is more difficult to obtain a smooth recovery of the groups (measured through global
ARI) than a local one (measured through averaged ARI), the former values  being globally smaller than the latter.  
In particular in the affiliation model, we observe that while the averaged ARI is rather good (all values close to
  $1$), the global one can be low (for e.g. with low group stability or medium group stability and $T=10$ time points). 
However in the identifiable cases,  we obtain  good performances for this global index (values above 0.8)
when group stability is not too low or when connectivity parameters are well enough separated (medium $\bb$ values). 
As expected, the clustering performances increase (i.e. ARI values increase) with group stability (from $\bpi$
  \emph{low} to \emph{high}) and with a better separation between the groups
connectivity behaviors (from \emph{low-} to \emph{medium+} easiness). 
When increasing the number of time points from $5$ to $10$, clustering indexes tend to be slightly larger, exhibiting a
smaller variance. However this is not always the case: for instance with low/medium group stability and $\beta=low+$, we observe
that the performances decrease from $5$ to $10$ time points (smaller ARI values). We
believe that this is due to the initialization of our procedure: with $T=10$ time points, it is more likely that the
groups membership differ from their initial value. As we use as a starting point a constant with time value for these
memberships, our algorithm is farther from the optimal value. 

Mean squared errors (MSE) for estimation of the transition parameter $\bpi$ are 
given in Supplementary Figure~\ref{fig:MSE}. We only show MSE for $\bpi$ as the MSE for $(\bb,\bg)$ are strongly correlated with the
clustering results.  This figure shows  that when groups are not
globally recovered, the MSE values may be high (up to 15\%).  However in most of the cases, these MSE are rather
small (less than 2\%) so that the dynamics of the groups membership is captured.

Now, we compare our results with other procedures. The models from~\cite{Yang_etal_ML11,Xu_Hero_IEEE} are the closest to our
setup. Since~\cite{Xu_Hero_IEEE} obtained comparable performances as the ones from~\cite{Yang_etal_ML11}, we focus on the latter
here. (In fact, \citeauthor{Xu_Hero_IEEE}'s method is faster, with slightly lower clustering performances
than~\citeauthor{Yang_etal_ML11}'s one.)
Thus, we use the offline version of the algorithm proposed in~\cite{Yang_etal_ML11} (Matlab code is available on the web
site of the first author). We ran their code on the same setup as above. When relying on default values of the algorithm,
the results obtained are very poor, with ARI values smaller than $10^{-2}$ in general (data not shown). We note that the authors
do not discuss initialization and simply propose to start with a  random partition of the nodes, which proves to be a
bad strategy. In order to make fair
comparisons, we thus decided to combine their algorithm with our initialization strategy. Results are presented in
Supplementary Figure~\ref{fig:Yang}. 

From this figure, we can see that putting appart our initialization strategy, our procedure
outperforms~\citeauthor{Yang_etal_ML11}'s one (they globally have much lower ARI). Indeed, their method obtains good performances only in a few cases:
$(\pi_{high},\beta\in \{$\emph{medium+; med w/ affiliation}$\}, T\in \{5,10\})$ ; $(\pi_{high},\beta\in\{low+,medium-,
T=5)$ and $(\pi_{medium}, \beta\in \{$\emph{medium+; med w/ affiliation}$\} ,T=5)$. 
In all these cases, we can see that the method's performances are due to a very good
initialization. Now, when the true classification is farther from initialization, the performances considerably drop. In
particular, for intermediate cases (e.g. medium group stability or high group stability with $T=10$), we can see that our method still succeeds in obtaining a good partition
(Figure~\ref{fig:ARI}) while this is not the case for~\citeauthor{Yang_etal_ML11}'s one (Supplementary Figure~\ref{fig:Yang}).

\subsection{Model selection}
\begin{figure}
  \centering
  \includegraphics[width=10cm,height=5cm]{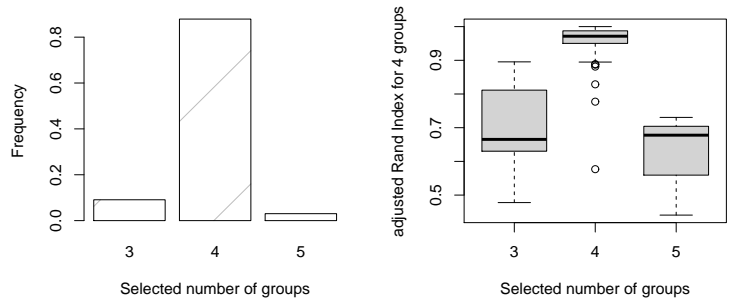}
  \caption{Estimation of the number of groups via ICL criterion. Left panel shows the frequency of the selected
    number of groups. Right panel shows ARI of the classification obtained with 4 groups depending on the
    selected number of groups.}
  \label{fig:ICL}
\end{figure}

We simulate a binary dynamic dataset with $Q=4$ groups, transition matrix
between states satisfies $\pi_{qq}=0.91$ and $\pi_{ql}=0.03$
for $q\neq l$. Bernoulli parameters are chosen as follows: we draw i.i.d. random variables $\{\epsilon_{ql}\}_{1\le
    q\le l\le 4} \in [-1,1]$ and then choose 
  \begin{align*}
\forall q\in \Q, \quad     \beta_{qq}&= 0.4+\epsilon_{qq}0.1 , \\
\forall q\neq l \in \Q^2, \quad \beta_{ql} & =0.1 +\epsilon_{ql}0.1 .
  \end{align*}
We generate 100 datasets under this model and estimate the number of groups relying on ICL criterion. Results
are presented in Figure~\ref{fig:ICL}. We observe that the correct number of groups is recovered in 88\% of the cases (left panel). Moreover, the right panel shows that when ICL selects only 3 groups, ARI of the
  classification with 4 groups is rather low (less than 80\%). This shows that in those cases, classification with 4
  groups is not the correct one, so that \texttt{VEM} algorithm seems responsible for bad results (optimum has not been reached) more than the
  penalization term.  

\section{Revealing social structure in dynamic contact networks}
\label{sec:real_data}
Dynamic network analysis has recently emerged as an efficient method for revealing social structure and organization in humans and animals.
Indeed, many studies are now beyond the analysis of static networks and take advantage of longitudinal data on the long term,
for instance during days or years of observations, that allow for constructing dynamic social networks.
In particular, contact networks built from field observations of association between animals or from sensors-based measurements, are now currently available in Ecology or Sociology.
In this section, we show that our statistical approach is a suitable tool to analyze  dynamic contact networks from the literature.

\subsection{Encounters among high school students}
Describing face-to-face contacts in a population (in our case, a classroom) can play an important role in 1/ understanding if there is a peculiar non-random mixing of individuals that would be a sign for a social organization and 2/ predicting how infectious diseases can spread, by studying the crosslink between the contacts dynamics and the disease dynamics. As a first stone, it is therefore mandatory to find an appropriate model to analyse these contacts and we propose to use our dynamic SBM to achieve this step.

The dataset  consists in face-to-face encounters of high school students (measured through the use of wearable sensors)
of a class from a French high school \cite[see][for a complete description of the experiment]{sociopat}.
In this class called 'PC' (as students focus on Physics and Chemistry), interactions were recorded during  4 days (Tuesday to Friday) in Dec. 2011. 
We kept only the 27 (out of 31) students that appear every day, i.e. that have at least one interaction with another student during each of the 4 days.  
Interaction times were aggregated by days to form a sequence of 4 different networks.
These are undirected and weighted networks, the weight of an interaction between
two  individuals being  the  number of  interactions  between these  2
individuals divided by the number of time points for which at least two individuals interacted; 
thus  a  non  negative  real  number that  we  call  {\it  interaction
  frequency}. After examination of  the distribution of these weights,
we   choose  to   discretize   these  data   into   $M=3$  bins   (see
 Example~\ref{ex:finite} in Section~\ref{sec:examplesM} from Supplementary
  Materials) corresponding to {\it low, medium} and {\it high} interaction frequency. 
As already mentioned, our  model selection criterion is not  fitted to this case (see Section~\ref{sec:examplesM} from Supplementary
  Materials for more details). We thus  choose to rely instead
on the 'elbow' method, applied to the complete data log-likelihood. It
consists in identifying a change of slope on the curve that represents
this complete data log-likelihood for different values of $Q$. 
The method selects $Q=4$ groups (see Supplementary Figure~\ref{fig:lkl}) and we now
present the results obtained with our model fitted with $Q=4$ groups.

\begin{figure}[h!]
  \centering
  \includegraphics[width=6cm,height=6cm]{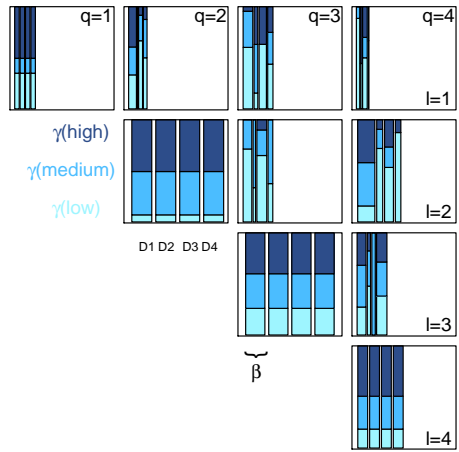}
  \caption{Summary  of the  interaction  parameters $\hat{\bb}$  and
    $\hat{\bg}$ estimated by our model with $Q=4$ grous on the dataset of
    interactions in the 'PC' class \citep{sociopat}. 
In each cell $(q,l)$ with $1\le q\le l\le 4$, there are $T=4$ barplots
corresponding to the  4 measurements (Tuesday to  Friday). Each barplot
represents the  distribution of the parameter  $\gamma_{ql}^t$ for the
three categories of interaction frequency  ({\it low, medium} and {\it
  high}). The width of each barplot  is proportional to the sparsity parameter $\beta_{ql}^t$.
We recall  that when  considering the  diagonal cells
$(q,q)$, parameters do not depend on $t$ anymore. }
  \label{fig:matrix}
\end{figure}

\begin{figure}[h!]
  \centering
  \includegraphics[trim=0 2.5cm 0 2.5cm,clip,width=7cm,height=6cm]{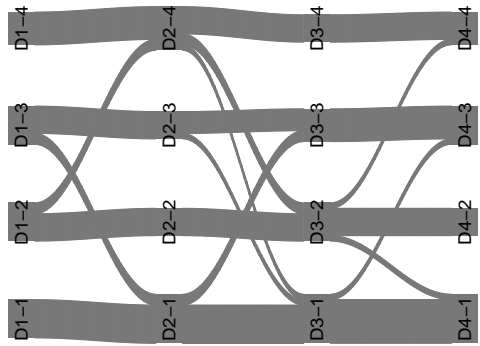}
  \caption{Alluvial plot showing the  dynamics of the group membership
    estimated by our model on the dataset of interactions in the 'PC' class~\citep{sociopat}. 
Each line is a flux that represents the move of one or more students from a group to another group ($Di-k$
indicates group $k$ for day $i$). 
The thickness of the lines is proportional to the number of students and the total height represents the 27 students.}
  \label{fig:alluvial}
\end{figure}

We  observe that  groups 2  and 3  are composed  by students  that are
likely to interact together (i.e. $\hat \beta_{22}$ and $\hat \beta_{33}$ are close to 1, see Figure~\ref{fig:matrix}). 
Furthermore, the  frequency of their interactions  inside their groups
is   higher    than   in    the   rest    of   the    network   ($\hat
\gamma_{qq}(low)<\hat                    \gamma_{qq}(medium)<\hat
\gamma_{qq}(high)$  for   $q=2,3$, same figure).   
These two groups form two communities such as defined in~\cite{Fortunato}. 
Moreover, we observe that both groups include a certain number of individuals
(3 and  4 respectively) that permanently  stay in the group  over time
(see Figure~\ref{fig:alluvial}).  These individuals may play  the role
of  'social  attractors' or  'core  leaders'  around which  the  other
students are likely to gravitate. 
Group 4 displays  a similar pattern of community  structure, with much
less interaction (intermediate value of  $\hat \beta_{44}$) but also a
significant level of interaction with group 2 (Figure~\ref{fig:matrix}). 
Interestingly, groups 2 and 4 also exchange students over time (see fluxes between groups in Figure~\ref{fig:alluvial}) and  this could reflect some cooperation or affinity between the students of these two groups.
Group 1 is quite stable over time (7 permanent members, see Figure~\ref{fig:alluvial}) and is characterized by a low
rate of interactions inside and outside the group (low $\hat \beta_{1q}^t$ values in Figure~\ref{fig:matrix}). 
It clearly gathers isolated students, but this does not mean that they
do not interact with any student, they usually do so, but with a small
number  of partners.  Therefore,  we  do not  only decipher  evolving
communities \citep[such as in][]{Yang_etal_ML11} but we also highlight the dynamics of aloneness inside this class.

We now investigate if gender differences may help in (a posteriori) explaining or refining the interaction patterns that we reveal.
We first note that group 3 is exclusively composed by male students (Supplementary Figure~\ref{fig:alluvial-gender}). This observation along with the previous conclusions suggests that group 3 may be a closed/exclusive male-community.
Meanwhile, some of these male students move to group 1 which is partly composed by a 'backbone' of female students that
stay in group 1 (Supplementary Figure~\ref{fig:alluvial-gender}).
Moreover, we clearly observe that female students are likely to stay in their group 
(most of the moves between groups are realized by males), 
and that a majority of them are in low-interacting groups 1 and 4.
But  not any  female student  moves  between these  two groups,  which
supports a  clear dichotomy  pattern in  the female  organization with
respect to male organization. 
In summary, we show  evidence for some gender homophily~\cite[see][for
a precise definition]{sociopat}, i.e. gender is a key factor for explaining the dynamics of the interactions between these young adults.

Lastly, we note that both information captured by our model (say $\bb$ and $\bg$) are often convergent/correlated in this case, 
but we note  that studying this network with a  binary model (i.e. not
considering  the  interaction frequency)  does  not  allow to  capture
interesting structure (data not shown). Therefore, the presence/absence of an interaction as well as its frequency are important and require an explicit modelling such as in our approach.


  \subsection{Social interactions between animals}
\label{sec:animals}  
Interactions among animals are dynamic processes. 
  How and why the topology of the network changes (or not) over time is of primary interest to understand animal societies.
  Here we analyze two datasets of animal contact networks, the first one dealing with migratory birds ({\it sparrows})
  and the second with  indian equids ({\it onagers}). Both datasets are analysed with the extended model presented in
  Section~\ref{sec:extensions} from Supplementary Materials. 
  
\begin{figure}[h!]
  \centering
  \includegraphics[width=6cm,height=6cm]{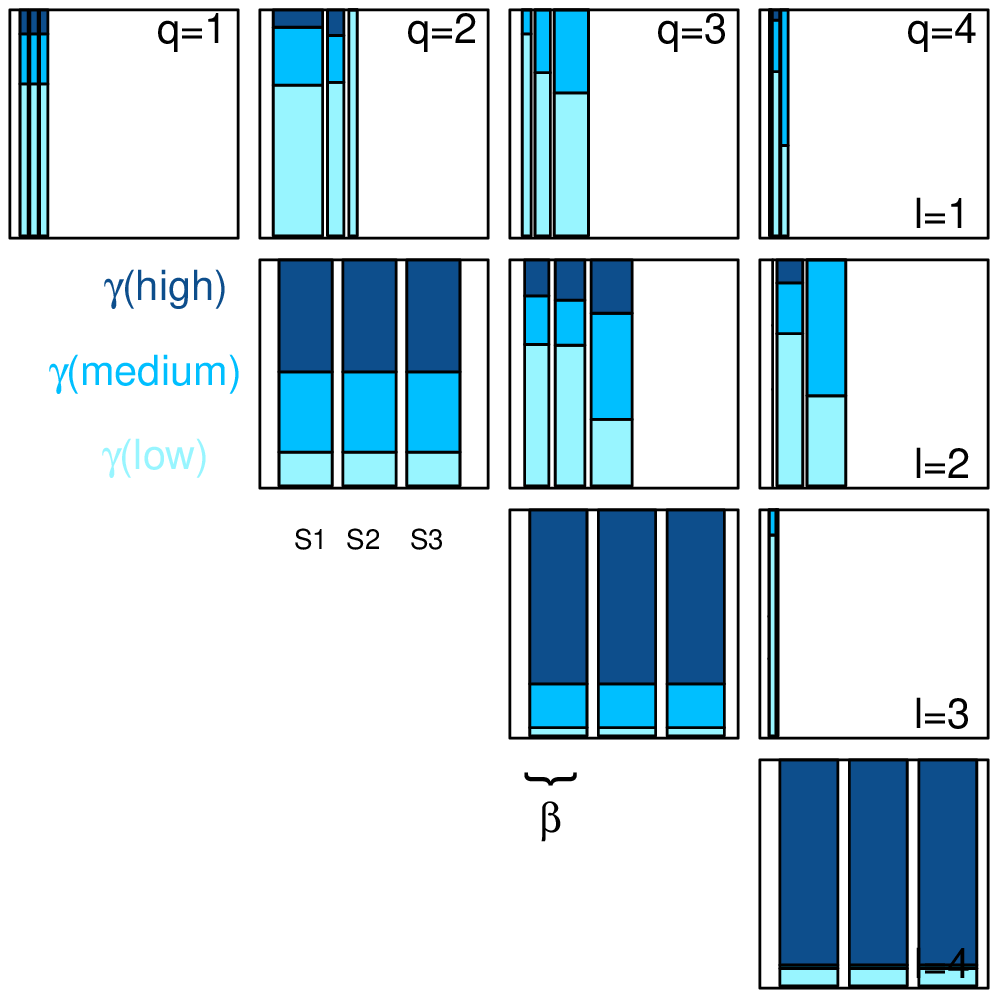}
  \includegraphics[width=6cm,height=6cm]{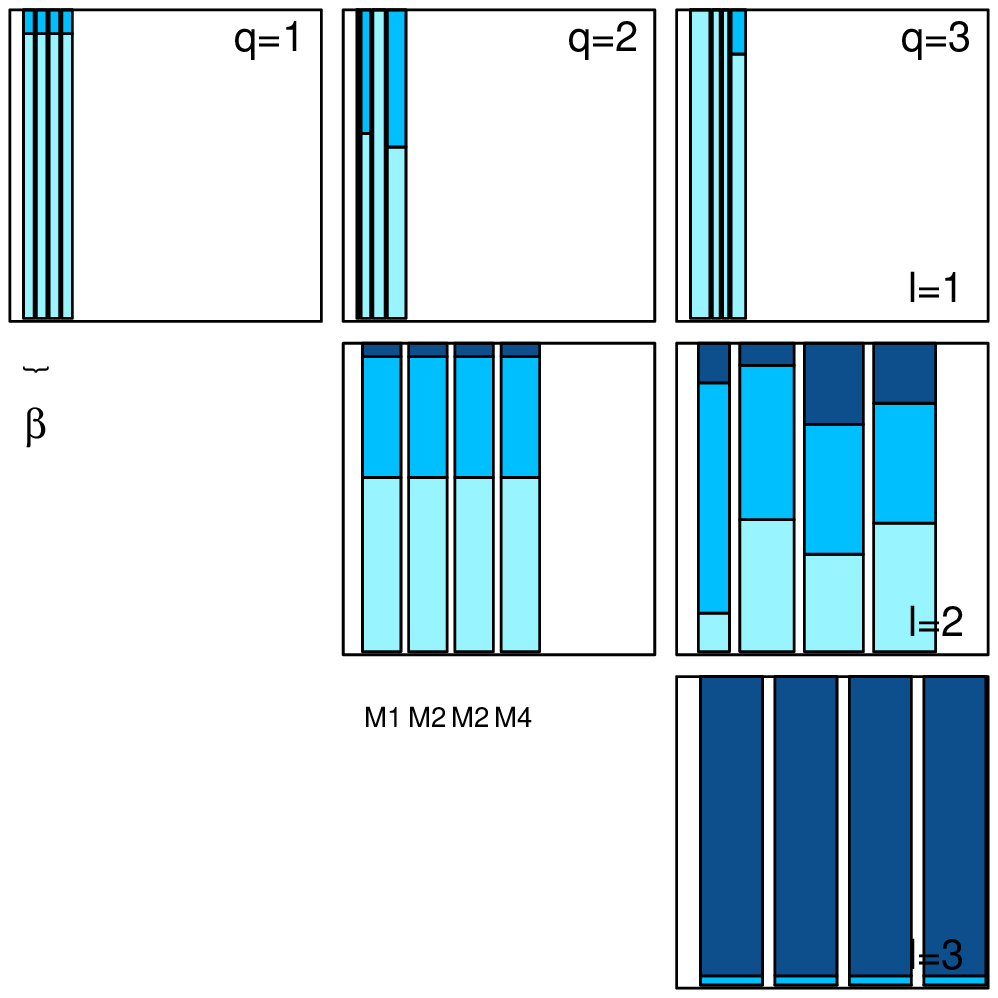}
  \caption{Summary  of the  interaction  parameters $\hat{\bb}$  and
    $\hat{\bg}$ estimated by our model with $Q=4$ groups on the dataset of sparrows~\cite[left panel,][]{sparrows} and $Q=3$
    groups on dataset of onagers~\cite[right panel,][]{onagers}, respectively. Same principle as in Figure~\ref{fig:matrix}.}
  \label{fig:matrixanimal}
\end{figure}

\begin{figure}[h!]
  \centering
  \includegraphics[trim=0.75cm 2cm 0.5cm 2cm,clip,width=6.5cm,height=6cm]{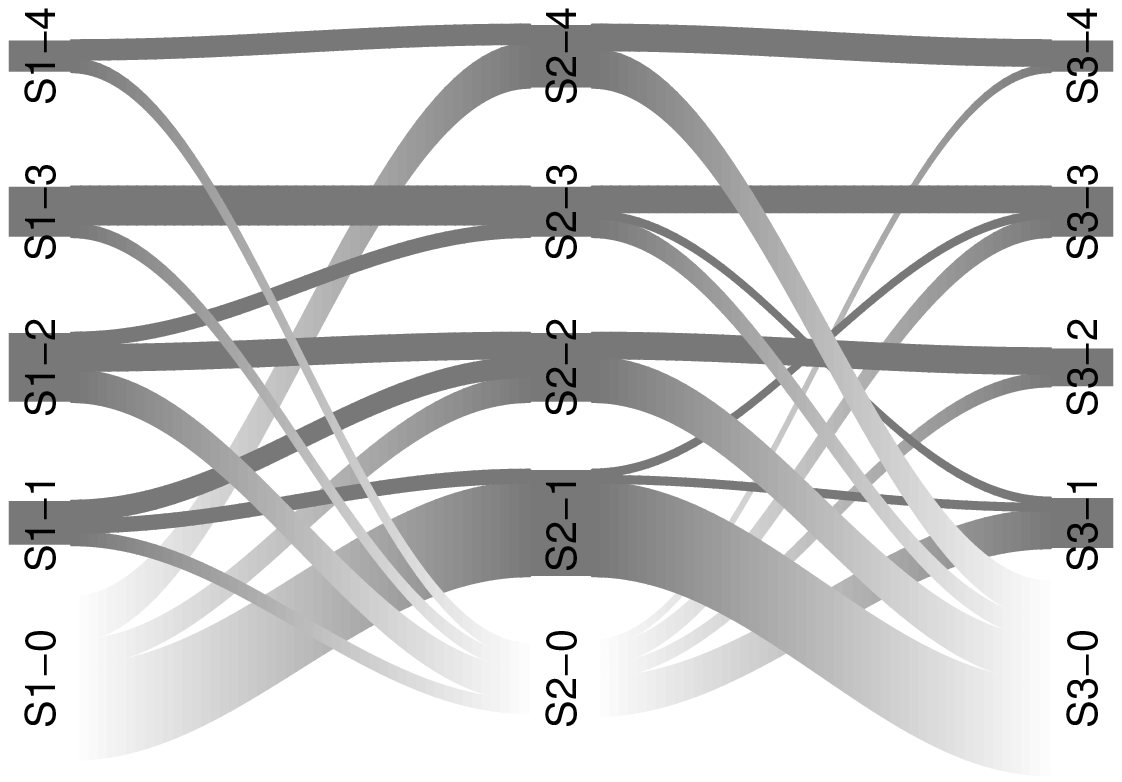}
  \includegraphics[trim=0.5cm 2cm 0.75cm 2cm,clip,width=6.5cm,height=6cm]{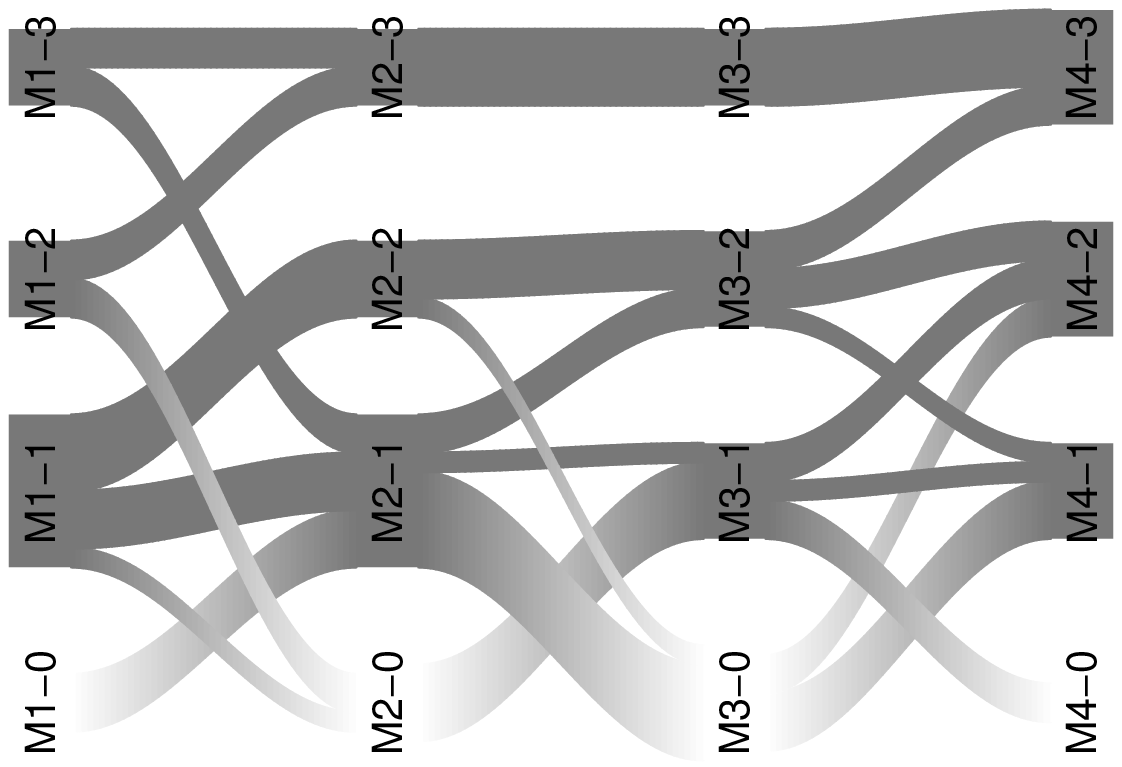}
  \caption{Alluvial plot showing the  dynamics of the group membership
    estimated by our model on the datasets of interactions between 69 sparrows (left,~\cite{sparrows}) and 23 onagers
    (right,~\cite{onagers}) respectively. Same principle as in Figure~\ref{fig:alluvial} (with $Si-k, Mi-k$ indicating
    group $k$ at Season or Month $i$). A fake group (group 0) gathers absent animals at a specific time step and fuzzy fluxes represent arrival/departure to/from a group from/to group 0.}
  \label{fig:alluvialanimal}
\end{figure}

  Sparrows were captured and marked during winters of 2010-12 in a small area \cite[The University of California,
  Santa Cruz Arboretum,][]{sparrows}. During these three seasons, Shizuka and colleagues  recorded birds interactions (into {\it flocks}, i.e. individuals in the same place at the same time) and they aggregated their observations by seasons.
  They observed 69 birds in total, but there was a significant turnover of birds due to mortality and recruitment and  only some of these 69 birds are present at each season (31, 46 and 27 birds, respectively). 
  The dataset is therefore composed by $T=3$ undirected and weighted networks, with specified presence/absence of nodes
  at each of the three time steps. Edges are weighted by the number of times  pairs of birds have been seen together at
  the same place and time (if zero, no edge). The authors  identified re-assembly of same communities (as
  defined previously) across seasons despite the birds turnover. This stability is due to social preferences across
  years between individuals that re-join the community located in the same area of the site \citep{sparrows}. Our model is a
  perfect candidate to fit these observations: indeed, constraints from Equation~\eqref{eq:intra_gp_const} are
  appropriate in this case where the communities keep existing over time (and therefore the parameters remain stable
  over time) but the membership is evolving (in particular, due to the presence/absence of birds in the three seasons).
  As previously, we discretized the edge weights into $M=3$ bins ({\it low, medium} and {\it high} interaction level)
  and we selected $Q=4$ groups with the 'elbow' method (data not shown). 
  Examination of the estimated parameters $\hat \bb$ and $\hat \bg$ (Figure~\ref{fig:matrixanimal}, left panel) reveals
  that groups 2, 3 and 4 are clear communities (with different intra-group behaviors) that eventually correspond to those revealed by Shizuka {\it et al}. Most
  of that, our method proposes to gather peripheral birds into group 1. Clearly, we observe some stability across
  years with individuals staying in communities 2, 3 and 4 over time  (see horizontal fluxes in
  Figure~\ref{fig:alluvialanimal}, left panel) and that are joined by incoming birds (see fluxes from the fake group 0
  of absent birds in this figure). All these observations confirm the analysis in \cite{sparrows} and demonstrate that
  our modelling approach is particularly suited to such datasets. 
  
  Onagers were observed in the Little Rann of Kutch, a desert in Gujarat, India \citep{onagers}.
  Each time a herd (group) has been encountered, association between each pair of individuals in the group was recorded.
  We retained the data association of 23 individuals present at least once between February and May 2003 and we
  aggregated interactions by month. The dataset contains therefore $T=4$ undirected and weighted networks, with specified presence/absence of nodes each month. Edges are weighted by the number of times  pairs of onagers belong to the same herd (if zero, no edge). 
  Again, we discretized the edge weights into $M=3$ bins  ({\it low, medium} and {\it high} interaction level) and we selected $Q=3$ groups. Visual inspection of the
  estimated parameters (Figure \ref{fig:matrixanimal}, right panel) shows that cluster 1 gathers peripheral
  onagers that can actually stay away from the others because predators have been extirpated from this habitat (and so,
  no collective protection strategy is required, \cite{onagers}). Cluster 2 is composed by {\it followers} onagers which
  have some interactions between them and much more with those of group 3 whereas onagers in group 3 form a {\it rich
    club} community  \cite[i.e. clique of hubs as defined in][]{rich}, with high values of estimates  $\hat \beta_{33},
  \hat \gamma_{33}(high)$. This community is evolving over time by integrating one or two onagers during the
  successive months. Interestingly, the social integration process is revealed and somehow hierarchical: previously
  absent onagers (fake group 0 in Figure~\ref{fig:alluvialanimal}, right panel) are likely to integrate group 1, onagers of group 1 can
  possibly move to the followers groups (i.e. group 2), and a few followers can be integrated over time in the central
  rich club community (group 3). Again, the structure of the onagers social network remains persistent over time
  \citep[see similar conclusions in][]{onagers} and our model is therefore particularly adapted and efficient in this case.	
    
\paragraph*{Acknowledgments}
We would like to thank Tianbao Yang for making his code available from his web page.
We sincerely acknowledge the SocioPatterns collaboration for providing
the high school dataset (http://www.sociopatterns.org) and especially 
Alain Barrat for interesting discussions.
This work was performed using the computing  facilities of the CC LBBE/PRABI.


\bibliographystyle{chicago}
\bibliography{dynSBM}


\newpage
\setcounter{page}{1}
\setcounter{section}{0}
\setcounter{figure}{0}
\pagenumbering{Roman}
\renewcommand\thesection{\Alph{section}}
\renewcommand\thefigure{S\arabic{figure}}
\pagestyle{plain}

\begin{center}
  {\Large
Supporting Materials for Statistical clustering of temporal networks through a dynamic 
  stochastic block model by C. Matias and V. Miele}
\end{center}
\vspace{1cm}


\section{Counter example of identifiability when groups memberships and connectivity parameters vary freely}
\label{ex:nonident}
Here, we exhibit an example where the parameters are non identifiable when both groups memberships and connectivity
parameters may vary across time without any constraint. 

  Let us consider the case of $Q=2$ groups and (for simplicity of notation) $2T$ time points. We fix 
a first parameter value $\theta=(\bpi,\bb,\bg)$ defined by
  $\bpi=Id$ the size-two identity matrix and $(\beta^t,\gamma^t)=(\beta,\gamma)$ are chosen constant with
  $t$. In the following, we let $\phi_{ql}(\cdot)$ denote the constant (with time) conditional density distribution of any
  $Y_{ij}^t$ given $Z_{iq}^tZ_{jl}^t=1$, under parameter value $\theta$. The latent process has
  stationary distribution $\ba=(1/2,1/2)$ and since the latent configuration is drawn at the first time point and
  stays constant ($\bpi$ is the identity), it can be seen that the distribution on the set of observations $\Y$ is given by 
  \begin{align*}
\pr(\Y) & = \frac 1 {2^N} \sum_{q_1\dots q_N \in \Q ^N }
\prod_{1\le i<j\le N} \prod_{t=1}^{2T} \phi(Y^t_{ij}; \beta^t_{q_iq_j},\gamma^t_{q_iq_j}) \\
& = \frac 1 {2^N} \sum_{q_1\dots q_N \in \Q ^N }
\prod_{1\le i<j\le N}\prod_{t=1}^{2T}  \phi_{q_iq_j}(Y^t_{ij}).
  \end{align*}
Now we consider a second   parameter value $\tilde \theta=(\tilde \bpi,\tilde \bb,\tilde \bg)$ such that 
\[
\tilde \bpi = 
\begin{pmatrix}
  0 & 1 \\
1 & 0 
\end{pmatrix}, 
\]
which corresponds to the same latent stationary distribution $\ba=(1/2,1/2)$ but now the latent configuration  is
drawn at the first time point and then each node switches group at each following time point. For any $q\in \{1,2\}$, we
let $\bar q$ denote the unique value such that $\{q,\bar q\}=\{1,2\}$.  
Moreover, for any $q\in \{1,2\}$, we set the within group parameter at time  $t=1$ to  
$(\tilde \beta^1_{qq},\tilde \gamma^1_{qq})=(\beta_{qq},\gamma_{qq})$, or equivalently, we set the conditional
distribution $\tilde \phi^1_{qq}$ of $Y_{ij}^1$ given $Z_{iq}^1Z_{jq}^1=1$, under parameter value $\tilde \theta$, equal
to previous value $\phi_{qq}$. Then, we switch the within group parameters values at each time point by setting  
\[
\forall t\ge 1, \quad 
 \tilde \phi^{t+1}_{11} =\tilde \phi^{t}_{22}    \text{ and }
\tilde \phi^{t+1}_{22} =\tilde \phi^{t}_{11}  . 
\]
Finally, the across group parameter is not modified through time and we set $\tilde \phi^{t}_{12}=\phi_{12}$. Now, we can write the
distribution of $\Y$ under parameter value $\tilde \theta$
  \begin{align*}
\mathbb{P}_{\tilde \theta}(\Y) 
& = \frac 1 {2^N} \sum_{q_1\dots q_N \in \Q ^N }
\prod_{1\le i<j\le N}\tilde \phi^1_{q_iq_j}(Y^1_{ij})\tilde \phi^2_{\bar q_i \bar q_j}(Y^2_{ij})\dots \tilde
\phi^{2T-1}_{q_iq_j}(Y^{2T-1}_{ij}) \tilde \phi^{2T}_{\bar q_i \bar q_j}(Y^{2T}_{ij}) \\
& = \frac 1 {2^N} \sum_{q_1\dots q_N \in \Q ^N }
\prod_{1\le i<j\le N}\phi_{q_iq_j}(Y^1_{ij})\phi_{q_iq_j}(Y^2_{ij})\dots \phi_{q_iq_j}(Y^{2T}_{ij}),
  \end{align*}
so that $\pr^Y=\mathbb{P}_{\tilde \theta}^Y$. To conclude, it suffices
to show that there is  no global permutation $\sigma\in \mathfrak{S}_Q$ such
that $\tilde \theta=\sigma(\theta)$. 
This can be seen from the fact  that for any $\sigma \in \mathfrak{S}_Q$, we
have $\sigma(\tilde \bpi)=\tilde \bpi \neq \bpi$. 
Thus the  two parameters  $\theta,\tilde \theta$ are  not equal  up to
label switching while they produce the same distribution on the
observations. 
It follows that the parameter $\theta$ is not identifiable up to label
switching.
Note that the SBM part of the parameter is recovered up to local label
switching   as   choosing   the  permutations   $\sigma_{2t}=Id$   and
$\sigma_{2t-1}=(1,2)$ (the transposition  in $\mathfrak{S}_2$) for any
$1\le t \le T$, we obtain that $\sigma_t(\beta^t,\gamma^t)=(\tilde \beta^t,\tilde \gamma^t)$.


\section{Non identifiability in affiliation case (planted partition)}
\label{sec:affil_ident}
Identifying  the whole parameters  from  a  binary  affiliation  SBM  is  a
difficult task, as may be seen for instance by the many different
but always partial results obtained by~\cite{AMR_JSPI}.  In  their
Corollary 7,  the authors  establish that  \emph{when group  proportions are
known}, the parameters $\beta_{\text{in}} (:= \beta_{qq}$ for all $q$)
and $\beta_{\text{out}}(:=\beta_{ql}$ for all $q\neq l$)  of a binary affiliation
static SBM are identifiable.  In the weighted affiliation case, all parameters $(\ba,\bb^t,\bg^t)$ of a (static) SBM may
be identified~\citep[Theorem 13 in][]{AMR_JSPI}.
Following the proof of Proposition~\ref{prop:ident}, we could identify $(\ba,\bb,\bg)$ in dynamic affiliation SBM under
natural assumptions. 
Now, without an additional constraint on the transition matrix
$\bpi$,    it   is    hopeless    to    identify   the    transition
parameters.  Indeed, as the groups play similar roles at each time
step,  label  switching  between  different  time  steps  is  free  to
occur  and $\bpi$  may  not  be identified  (note  that assuming  that
$\beta^t_{\text{in}}$ or $\gamma^t_{\text{in}}$ does not
depend on $t$ is of no help here).
This may be seen for instance from the example constructed in Section~\ref{ex:nonident} that remains  valid in the  affiliation case. 
In fact, identifying $\bpi$ in dynamic  affiliation SBM seems to be as
hard  as  identifying  the  group proportions  in  static binary   affiliation
SBM. While static  affiliation often relies on an  assumption of equal
group  proportions, there  is  no simple  parallel  situation for  the
transition matrix $\bpi$  in the dynamic case  (the trivial assumption
$\bpi=Id$ is  far too  constrained). 
Let us now give some intuition on why $\bpi$ is difficult to recover. 
For instance, following the proof of Proposition~\ref{prop:ident}
and looking  at the distribution of  $(Y^t_{ij},Y^{t+1}_{ij})$ enables
us  to  identify  a  mixing   distribution  with  four  components  as
follows. Let $\delta^t_{\text{in}}$ (resp. $\delta^t_{\text{out}}$) 
be    a    shorthand    for    the    Dirac    mass    at    parameter
$(\beta^t_{\text{in}},\gamma^t_{\text{in}})$
(resp.   $(\beta^t_{\text{out}},\gamma^t_{\text{out}})$).   From   the
distribution  of   $(Y^t_{ij},Y^{t+1}_{ij})$,  we  identify   the  four
following components
\begin{align*}
&\Big(\sum_{qq'}\alpha_q^2\pi_{qq'}^2 \Big)\delta^t_{\text{in}}\otimes
\delta^{t+1}_{\text{in}} 
; \Big (\sum_{q}\sum_{l\neq m}\alpha_q^2\pi_{ql}\pi_{qm}\Big)\delta^t_{\text{in}}\otimes
\delta^{t+1}_{\text{out}} 
; \Big (\sum_{q\neq l} \sum_m \alpha_q\alpha_l\pi_{qm}\pi_{lm}\Big)\delta^t_{\text{out}}\otimes
\delta^{t+1}_{\text{in}} ; \\
&    \Big    (\sum_{q\neq    l}    \sum_{q'\neq    l'}\alpha_q\alpha_l
  \pi_{qq'}\pi_{ll'}\Big)\delta^t_{\text{out}}\otimes \delta^{t+1}_{\text{out}}. 
\end{align*}
Now relying on the knowledge of the proportions of each of these four components, it
can be seen that it is not easy to identify the individual values of $\bpi$.
Without a proper identification of  the transition matrix $\bpi$, we do
not  recover  the behavior  of  the  group membership  through  time.
Empirical  evidence for  label switching  between time  steps in
  the affiliation setup is given in Section~\ref{sec:experiments} from the Main Manuscript.


\section{Optimization with respect to $\tau(i,q)$}
\label{sec:annexe}
In this section, we provide the exact fixed point equation satisfied by the values $\hat \tau(i,q)$ maximizing
$J(\theta,\tau)$. We have 
\begin{align*}
\hat \tau(i,q) \propto &\; \alpha_q \prod_{j, j\neq i} \prod_{l=1}^Q
  [\phi^1_{ql}(Y^1_{ij})]^{\hat \tau(j,l)} \prod_{t\ge 2} \prod_{q_2\dots q_t}
  \Big(\frac{\pi_{q_{t-1}q_t}}{\hat \tau(t,i,q_{t-1},q_t)}\Big)^{\hat \tau(2,i,q,q_2)\dots\hat \tau(t,i,q_{t-1},q_t)} \\
& \times \prod_{t\ge 2} \prod_{q_2\dots q_t,l}  \phi^t_{q_t l}(Y^t_{ij})^{\htaum(t,j,l)\hat \tau(2,i,q,q_2)\dots \hat \tau(t,i,q_{t-1},q_t)} ,
\end{align*}
with the convention: whenever $t=2$ then $q_{t-1}=q$.
This equation is to be compared with our approximation given by~\eqref{eq:approx}.


\section{Estimation of $\bg$ and model selection: specific examples}
\label{sec:examplesM}
The  \texttt{M}-step equations concerning $\bg$
differ  depending on  the  specific choice  of  the parametric  family
$\{f(\cdot,\gamma),\gamma \in \Gamma\}$. We provide here many examples
of classical choices for these  parametric families. Remember that the
resulting conditional  distribution on  the observations is  a mixture
between an element from this family and the Dirac mass at zero. 
We  also provide  expressions  for ICL  criterion  in these  different
setups.

\begin{example}[Bernoulli]
  This specific case corresponds to  a degenerate family with only one
  element, the Dirac mass  at 1, namely $F(y,\gamma)=\delta_1(y)$. The
  parameter $\theta$ reduces to $(\bpi,\bb)$ for which updating expressions at the
  \texttt{M}-step have  already been given (see Proposition~\ref{prop:VEM}).  
Note that we  imposed the
  constraint  $\beta^t_{qq}$ constant  with  respect to  $t$, for  any
  $q\in \Q$.
Now, model selection is performed through~\eqref{eq:ICL_gen} where 
\[
pen(N,T,\bb,\bg)     =     pen(N,T,\bb)     =     \frac     1     2     Q
\log\Big(\frac{N(N-1)T}{2}\Big) + \frac     1     2     \frac{Q(Q-1)}{2}T\log\Big(\frac{N(N-1)}{2}\Big) .
\]
\end{example}


\begin{example}[Finite space]
\label{ex:finite}
Let us consider a finite set of $M\ge 2$ known values $\{a_1,\dots, a_M\}$ not
containing $0$ and 
\[
f(y,\gamma) = \sum_{m=1}^M \gamma(m) 1_{y=a_m},
\]
with $\gamma(m)\ge 0$ and $\sum_m \gamma(m)=1$. 
The value $\hat \bg$ that maximizes $J(\theta,\tau)$ with respect to $\bg$ is given by 
\begin{align*}
  \forall t, \forall q\neq l \in \Q^2, \forall m, \quad \hat \gamma^t_{ql}(m) &= \frac {\sum_{1\le i\neq j\le N}
    \taum(t,i,q)\taum(t,j,l)1_{Y^t_{ij}=a_m}}{\sum_{m,i,j}\taum(t,i,q) \taum(t,j,l) 1_{Y^t_{ij}=a_m}} ,\\
\forall q \in \Q, \forall m, \quad \hat \gamma_{qq}(m) &= \frac {\sum_{t=1}^T\sum_{1\le i\neq j\le N}
    \taum(t,i,q) \taum(t,j,q) 1_{Y^t_{ij}=a_m} }{\sum_{m,t,i,j}\taum(t,i,q) \taum(t,j,q) 1_{Y^t_{ij}= a_m}}    .
\end{align*}
These equations  remain valid when  considering a set of disjoint bins $\{I_m\}_m$  instead of
pointwise values $\{a_m\}_m$. 

In this setup, we do not propose a model selection
criterion  for  selecting  the  number  of  groups  $Q$.  Indeed,  our
investigations show  that a competition  occurs between the  number of
bins $M$ and  the number of groups  $Q$, so that in general  we end up
selecting only  $Q=2$ groups because  of a large number  of parameters
(data  not shown).   In fact,  this finite  distribution setup  may be
viewed as a  nonparametric model for which BIC-like  criterion (ICL is
of that type) are not suited. 
Section~\ref{sec:real_data} from the Main Manuscript proposes   another
approach to handle this case, relying on the 'elbow' method applied on the
complete data log-likelihood.  
\end{example}


\begin{example}[Poisson]
We consider the truncated Poisson distribution 
\[
f(y,\gamma) = (e^{\gamma}-1)^{-1}\frac{\gamma^y }{y !}  , \quad y \in \N\setminus\{0\},
\]
resulting in either a $0$-inflated or $0$-deflated Poisson when  mixed with the Dirac mass at
$0$. Let 
\[
\forall x > 0, \quad \psi(x) = \frac {x e^x} {e^x-1},
\]
which is a strictly increasing function and as such admits a unique inverse function $\psi^{(-1)}$ on
$(1,+\infty)$. Note that  $\psi^{(-1)}$ has no simple analytic expression but for any fixed $y\in (1,+\infty)$, the value $x=\psi^{(-1)}(y)$ may be
  found numerically by solving for $x$ the equation $x e^x/( e^x-1)-y=0$. 
Now the value $\hat \bg$ that maximizes $J(\theta,\tau)$ with respect to $\bg$ is given by 
\begin{align*}
  \forall t, \forall q\neq l \in \Q^2, \quad \hat \gamma^t_{ql} &= \psi^{(-1)} \Big(\frac {\sum_{1\le i\neq j\le N}
    \taum(t,i,q) \taum(t,j,l) Y^t_{ij}}{\sum_{i,j}\taum(t,i,q) \taum(t,j,l) 1_{Y^t_{ij}\neq
      0}} \Big),\\
\forall q \in \Q, \quad  \hat \gamma_{qq} &= \psi^{(-1)}\Big(\frac {\sum_{t=1}^T\sum_{1\le i\neq j\le N}
    \taum(t,i,q) \taum(t,j,q) Y^t_{ij}}{\sum_{t,i,j}\taum(t,i,q) \taum(t,j,q) 1_{Y^t_{ij}\neq
      0}} \Big).
\end{align*}
Model selection is obtained by maximizing~\eqref{eq:ICL_gen} with 
\begin{align*}
pen(N,T,\bb,\bg) = &\frac 1 2 \Big( |\{\beta_{qq}, q\in \Q\}| +Q \Big)
\log\Big(\frac{N(N-1)T}{2}\Big) \\
&+ \frac 1  2 \Big( |\{\beta^t_{ql}, 1\le
q<l\le Q, 1\le t \le T\}| +\frac{Q(Q-1)}{2}T\Big) \log\Big(\frac{N(N-1)}{2}\Big) .
\end{align*}
Moreover,  if  $\beta^t_{ql}=\beta_{\text{out}}$  does  not
depend on $t$ and $\beta_{qq}=\beta_{\text{in}}$, the penalty term in ICL becomes 
\begin{align*}
pen(N,T,\bb,\bg) = \frac 1 2 \Big(2+Q \Big)
\log\Big(\frac{N(N-1)T}{2}\Big) +\frac 1  2 \Big( \frac{Q(Q-1)}{2}T\Big) \log\Big(\frac{N(N-1)}{2}\Big) .
\end{align*}
\end{example}


\begin{example}[Gaussian homoscedastic]
\label{ex:Gauss}
Let us consider the Gaussian distribution 
\[
f(y,\gamma) = \frac 1 {\sqrt{2\pi}\sigma} \exp\Big( - \frac {(y-\mu)^2}{2\sigma^2}\Big), 
\]
where $\gamma= (\mu,\sigma^2) \in \R\times (0,+\infty)$. For parsimony reasons, we choose to consider the homoscedastic case where
the variance is constant across groups and simply denoted by $\sigma_t^2$. The value $\hat \bg$ that maximizes $J(\theta,\tau)$ with respect to $\bg$ is given by 
\begin{align*}
  \forall t, \forall q\neq l \in \Q^2, \quad \hat \mu^t_{ql} &= \frac {\sum_{1\le i\neq j\le N}
    \taum(t,i,q) \taum(t,j,l) Y^t_{ij}}{\sum_{i,j}\taum(t,i,q) \taum(t,j,l) 1_{Y^t_{ij}\neq
      0}} ,\\
\forall q \in \Q, \quad \hat \mu_{qq} &= \frac {\sum_{t=1}^T\sum_{1\le i\neq j\le N}
    \taum(t,i,q)\taum(t,j,q)Y^t_{ij}}{\sum_{t,i,j}\taum(t,i,q)\taum(t,j,q)1_{Y^t_{ij}\neq
      0}} ,\\
  \text{and } \forall t, \quad \hat \sigma_t^2& = \frac {\sum_{1\le i< j\le N} \sum_{1\le q,l\le Q}
 \taum(t,i,q)\taum(t,j,l) [Y^t_{ij}-\hat \mu^t_{ql}]^21_{Y^t_{ij}\neq 0} }{\sum_{i,j,q,l}\taum(t,i,q)\taum(t,j,l) 1_{Y^t_{ij}\neq 0}} .
\end{align*}
Here the remaining penalty term in~\eqref{eq:ICL_gen} for ICL criterion writes 
\begin{align*}
pen(N,T,\bb,\bg)&=                      \frac            1           2
                ( 2Q)\log\Big(\frac{N(N-1)T}{2}\Big) +\frac 1                2
                \Big( 2\frac{Q(Q-1)}{2}T\Big)\log\Big(\frac{N(N-1)}{2}\Big)\\
&=  Q \log\Big(\frac{N(N-1)T}{2}\Big) + \frac{Q(Q-1)}{2}T \log\Big(\frac{N(N-1)}{2}\Big).
\end{align*}
\end{example}

\section{Extension to varying number of nodes}
\label{sec:extensions}
In the present work, we limited ourselves to the case where the list of nodes $\{1,\dots,N\}$ stays constant across
time. However in real data applications it may happen that some actors enter or leave the study during the analysis. This may be
handled in a simple way as follows. Let us consider $V=\{1,\dots,N\}$ as the total list of individuals and for each time
step $t$, a subset $V^t$ of $V$ with cardinality $N_t$ of actors are present. Data is formed by a series of adjacency matrices
$\Y=(Y^t)_{1\le t\le T}$ where each $Y^t$ still  has size $N\times N$. For all pair of present nodes $i,j\in V^t$, entry $Y^t_{ij}$ characterizes the binary or weighted interaction between $i,j$ while for any $i,j\in V$
such that $i\notin V^t$, entry $Y^t_{ij}$ is set to $0$.
Now, we construct the latent process $\Z=(Z^t_i)_{1\le t\le T, i \in V}$ on an extended set $\Q_a=\Q\cup\{a\}$ where the
extra value $a$ stands for \emph{absent}. For each time step $t$ and  whenever $i\in V^t$, random variable $Z^t_i$ is constrained to vary in $\Q$ while for any $i\notin V^t$ we fix
$Z^t_i=a$. As previously, the random time series $(Z_i)_{i\in V}$ are supposed to be independent while for each
individual $i\in V$, the sequence $Z_i=(Z^t_i)_{1\le t\le T}$ forms an \emph{inhomogeneous} Markov chain with values in
$\Q_a$ and transitions $\bpi^t$ constrained by, for all $q,q'\in \Q$,
\begin{align*}
  \pi^t_{qa} &= \mathbb{P}(Z^t_i=a|Z^{t-1}_i=q) =1\{i\notin V^t\} , \\
\pi^t_{aq} &=  \mathbb{P}(Z^t_i=q|Z^{t-1}_i=a) = \alpha_q 1\{ i \in  V^t\}, \\
\pi^t_{qq'} &= \mathbb{P}(Z^t_i=q'|Z^{t-1}_i=q) = \pi_{qq'} 1\{ i\in V^t\}.
\end{align*}
Here, $\bpi=(\pi_{qq'})_{1\le q,q'\le Q}$ stands as previously for a transition matrix on $\Q$ of an irreducible aperiodic stationary Markov chain with stationary
distribution $\ba$. Note that the whole chain $Z_i$ is not stationary anymore. The probability of any trajectory of the
latent process simply writes  as
\begin{equation*}
  \mathbb{P}(\Z) = \prod_{i=1}^N \mathbb{P}(Z^1_i)\prod_{t=2}^T\mathbb{P}(Z^t_i|Z^{t-1}_i) = \prod_{q\in \Q}
  \alpha_q^{N_q} \times \prod_{q,q'\in \Q} \pi_{qq'}^{N_{qq'}} , 
\end{equation*}
where 
\begin{align*}
N_q& =\sum_{i\in V^1} 1\{Z^1_i=q\} + \sum_{t=2}^T\sum_{i\in V^t, i \notin V^{t-1}} 1\{Z^t_i=q\} ,\\
\text{and} \quad  N_{qq'}&=\sum_{t=2}^T \sum_{i\in V^{t-1}\cap V^t} 1\{Z^{t-1}_i=q, Z^t_i=q'\}.  
\end{align*}
As such, a node that would not
be present at each time point contributes to the likelihood only through the part of the trajectory where it is
present. 
Moreover, given the latent groups $\Z$, for any $i,j\in V^t$, the conditional distribution of $Y^t_{ij}$ is still given by~\eqref{eq:emission}
while whenever $i\notin V^t, j\in V$, we have $Y^t_{ij}$ is deterministic and set to $0$. Thus, a node absent at time
$t$ does not contribute to the likelihood of the observations. Generalization of our \texttt{VEM} algorithm easily
follows.

\section{Supplementary figures}
\label{sec:suppfigure1}
\begin{figure}[h!]
	\centering
	\includegraphics[width=8cm,height=7cm]{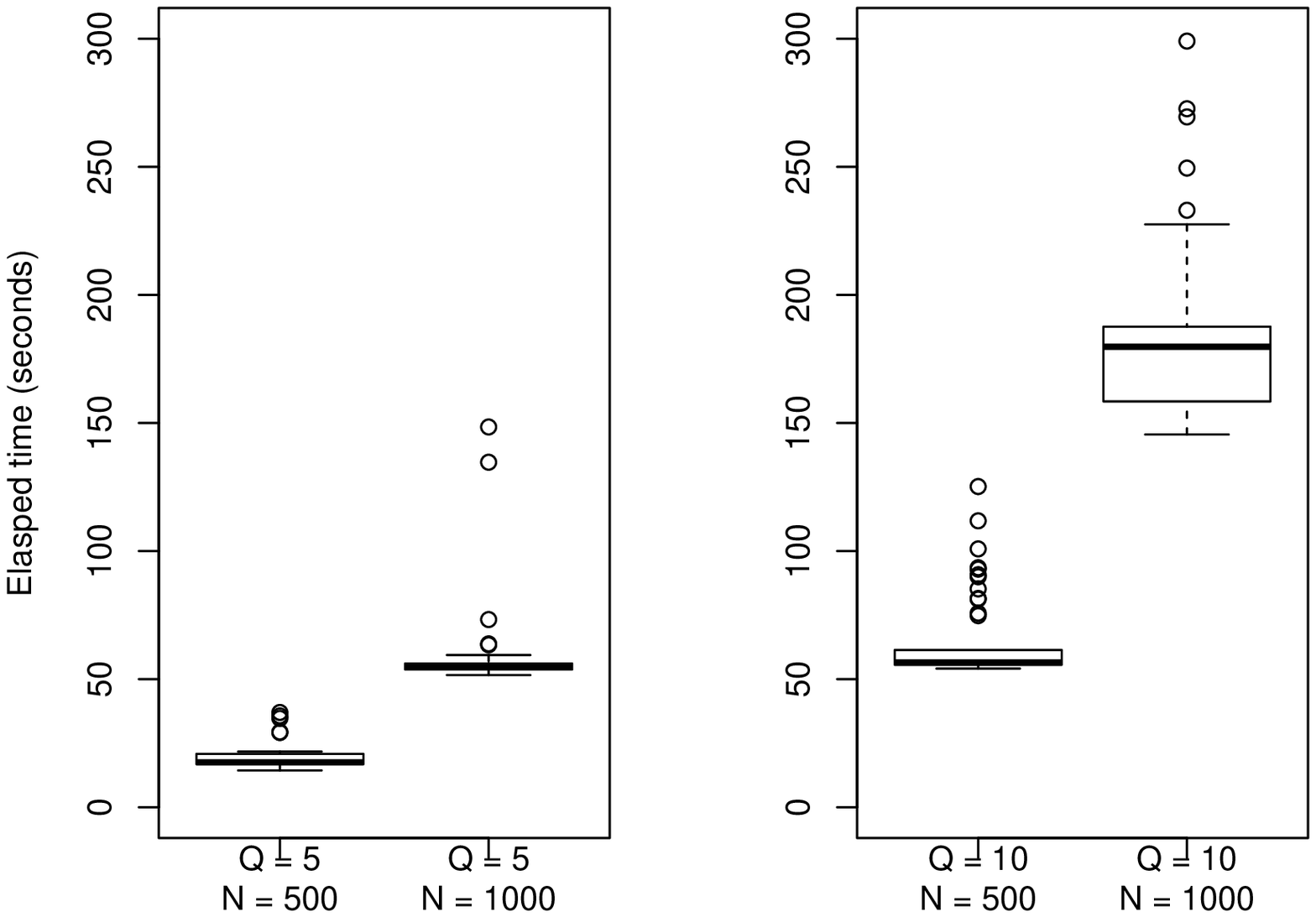}
	\caption{Boxplots of elapsed time (in seconds) for the estimation algorithm (including initialization) on 50 simulated datasets
          in the Bernoulli case with $\beta = medium+$ and high group stability (see Section~\ref{sec:experiments} in
          Main Manuscript), for $N=500,1000$ and $Q=5,10$. Performed on Intel Xeon E5-1620 v2 (8 threads, 3.70GHz). The
          stopping criteria is set to a relative difference on $J$ values of  \num{1e-4}.}
	\label{fig:cputime}
\end{figure}

\begin{figure}
 \centering
  \includegraphics[width=\textwidth,height=6cm]{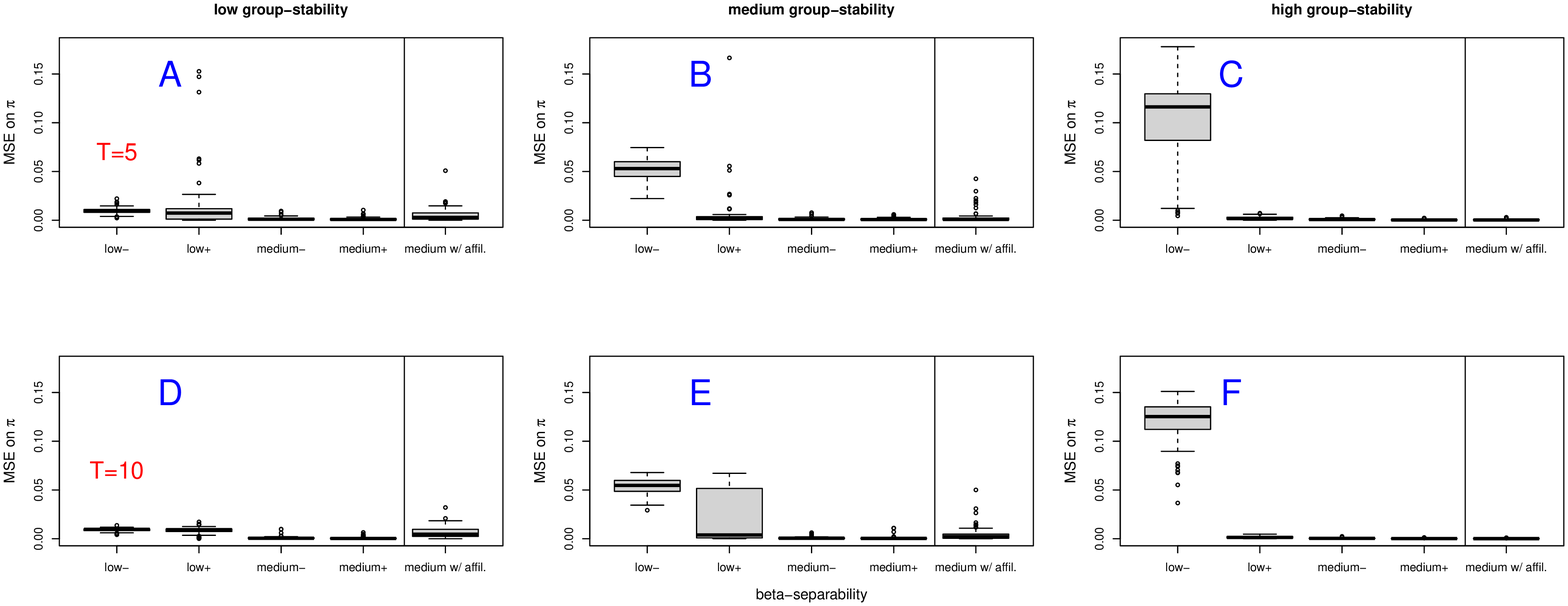}
   \caption{MSE boxplots for estimation of transition matrix $\bpi$ in different setups. From left to right: the
    three panels  correspond to $\bpi =  \bpi_{low}$ (panels A,D), $\bpi_{medium}$ (panels B,E) and
    $\bpi_{high}$ (panels C,F), respectively.  In each  panel, from left  to right:
    results  corresponding  to  $\bb=low-,low+,medium-,medium+$
    and affiliation case, respectively. First row: $T=5$ time points, second row: $T=10$.}
\label{fig:MSE} 
 \end{figure}

\begin{figure}
 \centering
  \includegraphics[width=\textwidth,height=6cm]{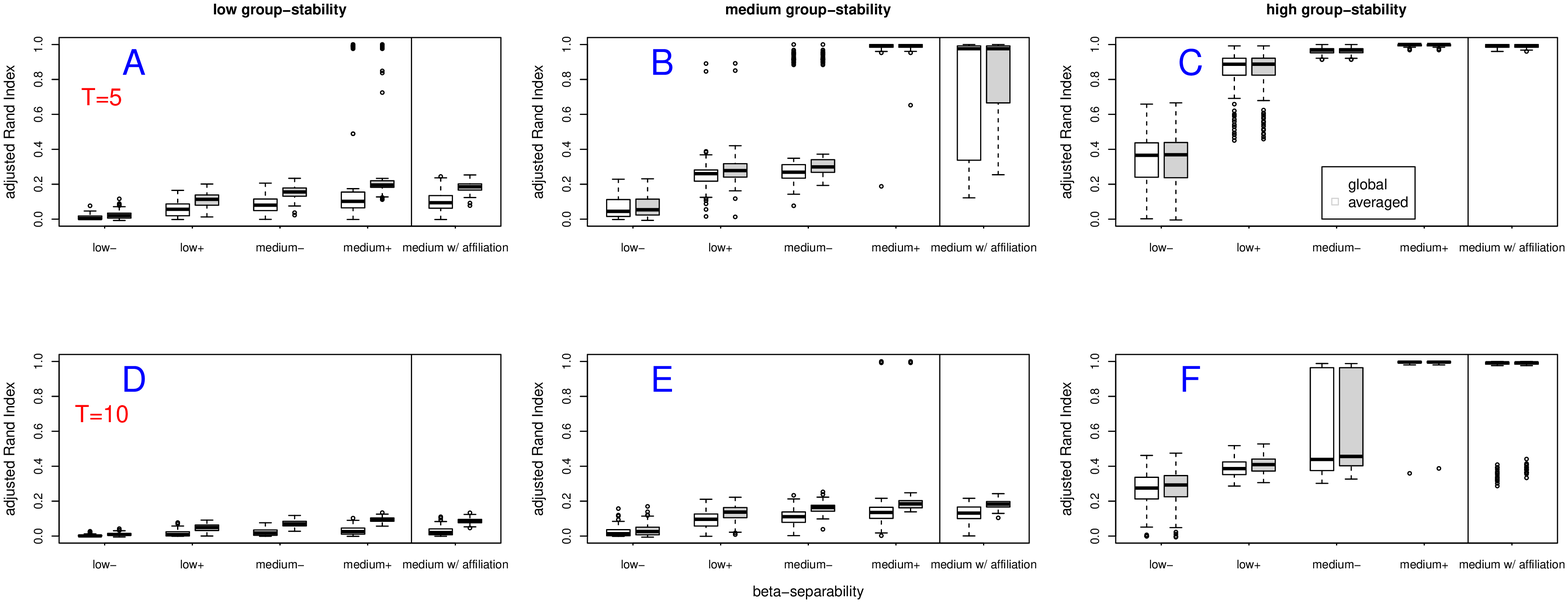}
  \caption{Boxplots  of global  ARI  (white, left)  and averaged  ARI
    (grey, right) in different setups for the combination of our initialization strategy with \citeauthor{Yang_etal_ML11}'s algorithm. From left to right: the
    three panels  correspond to $\bpi =  \bpi_{low}$ (panels A,D), $\bpi_{medium}$ (panels B,E) and
    $\bpi_{high}$ (panels C,F), respectively.  In each  panel, from left  to right:
    results  corresponding  to  $\bb=low-,low+,medium-,medium+$
    and affiliation case, respectively. First row: $T=5$ time points, second row: $T=10$.}
 \label{fig:Yang} 
 \end{figure}

\begin{figure}[h!]
  \centering
  \includegraphics[width=6cm,height=6cm]{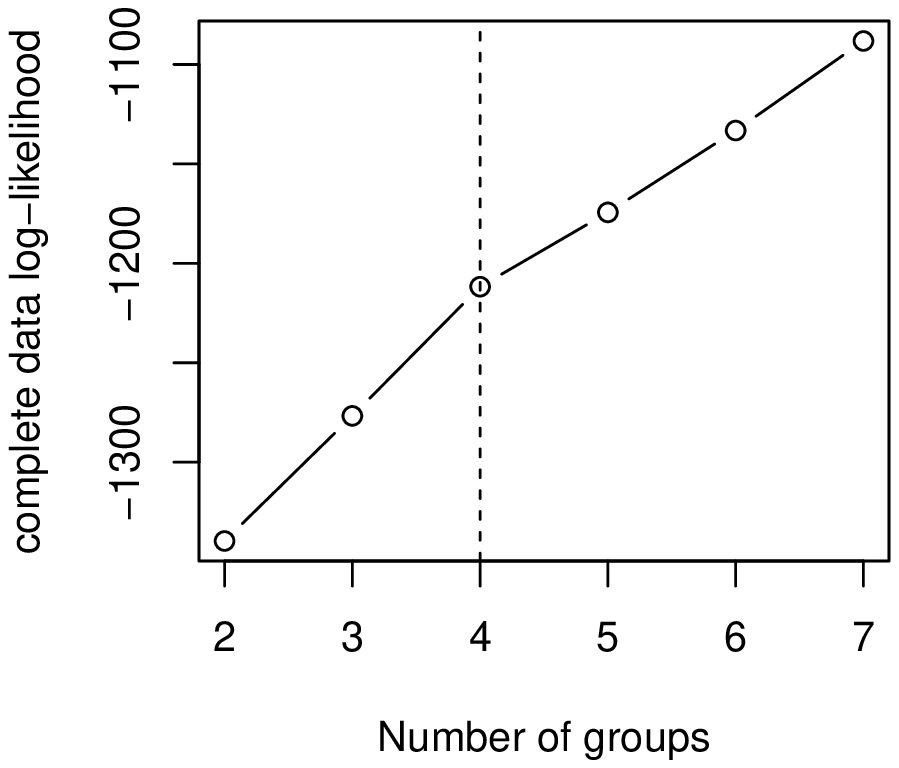}
  \caption{Complete  data   log-likelihood  estimated   for  different
    numbers of groups on the dataset of interactions in the 'PC' class~\citep{sociopat}. }
  \label{fig:lkl}
\end{figure}

\begin{figure}[h!]
  \centering
  \includegraphics[trim=0 2cm 0 2cm,clip,width=6.5cm,height=6cm]{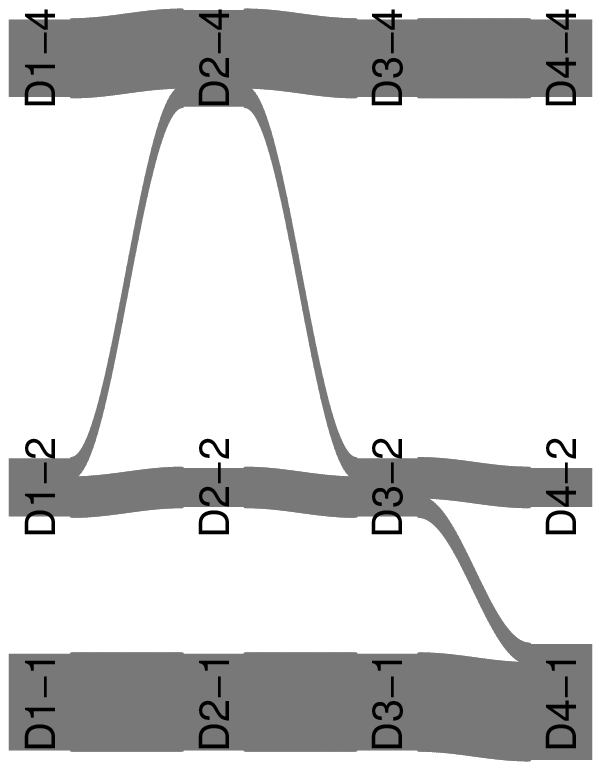}
  \includegraphics[trim=0 2cm 0 2cm,clip,width=6.5cm,height=6cm]{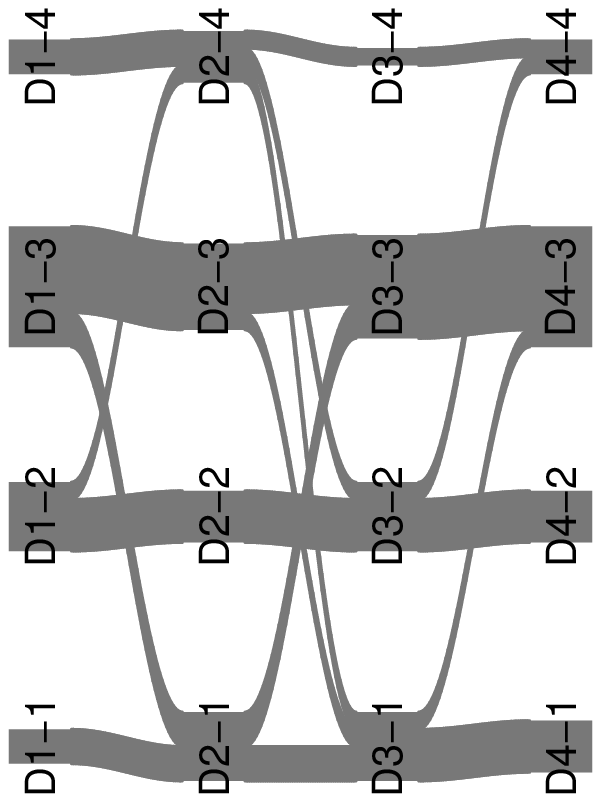}
  \caption{Same as in Figure~\ref{fig:alluvial} from Main Manuscript for the 12 female students (left panel) and the 15
    male students (right panel). }
  \label{fig:alluvial-gender}
\end{figure}

\end{document}